\documentclass[aps,pra,floatfix,twocolumn]{revtex4-1}
\usepackage{mathrsfs}
\usepackage{float}
\usepackage{graphicx}
\usepackage[english]{babel}
\usepackage{amsmath}
\usepackage{amssymb}
\usepackage{slashed}
\usepackage{bm}
\allowdisplaybreaks
\begin{document}


\title{Berry phase, entanglement entropy, and algebraic properties of ground states of BCS and BEC superfluids}
\author{Hao Guo$^1$, Yan He$^2$, and Chih-Chun Chien$^{3\ast}$}

\affiliation{$^1$Department of Physics, Southeast University, Jiulonghu Campus, Nanjing 211189, China}
\affiliation{$^2$College of Physical Science and Technology, Sichuan University, Chengdu, Sichuan 610064, China}
\affiliation{$^3$School of Natural Sciences, University of California, Merced, CA 95343, USA}
\email{cchien5@ucmerced.edu}

\begin{abstract}
By using Bogoliubov transformations to construct the ground states of fermionic Bardeen-Cooper-Schrieffer (BCS) superfluids and weakly-interacting Bose gases supporting Bose Einstein Condensation (BEC), their algebraic structures and implications can be analyzed in detail. Both ground states are generalized squeezed coherent states saturating a generalized Heisenberg uncertainty relation, and they acquire quantized Berry phases when the corresponding systems are transported along a closed path in their parameter spaces. While the Berry phase of the BCS ground state depends on the total particle number, the Berry phase of the BEC ground state depends only on the particles outside the BEC. The Berry phases are associated with magnetic monopoles in the parameter spaces and we found that the Dirac quantization condition is satisfied. Moreover, both ground states are entangled states of the fermion or boson quanta and we found the entanglement entropy quantifying the internal correlations. A fixed particle-number approach of fermionic superfluids does not saturate the generalized uncertainty relation, exhibits internal entanglement, and gives corresponding Berry phase. In addition, the algebraic structures of the ground states can be classified by the $q$-deformed Hopf algebra, $\bigoplus_{\mathbf{k}}h_{q_{\mathbf{k}}}(1)$ for bosons and $q$-deformed Hopf superalgebra $\bigoplus_{\mathbf{k}}h_{q_{\mathbf{k}}}(1|1)$ for fermions, respectively.
\end{abstract}

\maketitle

\section{introduction}
When Bardeen, Cooper and Schrieffer first formulated the BCS theory of superconductors, the ground state was constructed as a collection of Cooper pairs~\cite{BCS57}. Later on, the BCS ground state was shown to be related to coherent states~\cite{CTP02,HuangPLA08}. By analyzing the algebraic structure of the BCS ground state, we will show that it is actually a second order nilpotent SU(2) generalized coherent state. The structure of the BCS ground state indicates that, instead of being constructed from an assembly of Cooper pairs, it can be directly derived by implementing a Bogoliubov transformation of fermions. The BCS ground state and its generalized form with a parametrized order parameter are shown to explicitly satisfy the generalized uncertainty relation.

The BCS ground state is a collection of different numbers of Cooper pairs and is constructed in grand canonical ensemble~\cite{MahanBook}. Ground states with fixed particle number have been contemplated after the BCS theory, and it was found that in the thermodynamic limit, the BCS theory and fixed particle-number approach should lead to the same equations of state (see Ref.~\cite{BarangerPR63} and references therein). The fixed particle-number approach has been discussed by different authors in the literature~\cite{Peierls91,Peierls92,Leggett01,LeggettBook,Koizumi15}. We will show that, while the two approaches (BCS theory and fixed particle-number wave function) agree on things like the equations of state, internal entanglement, and Berry phase, the generalized uncertainty relation singles out the coherent-state structure of the BCS wave function. While the BCS ground state may not apply to unconventional superconductors~\cite{LeggettBook,SCBook}, here we focus on its algebraic structure and implications.

The entanglement entropy measures the correlation between two parts of a bipartite system \cite{QP_book}.
The Bogoliubov transformation leading to the generalized BCS ground state reveals entanglement between the two components (or spins) of the fermions, which is not transparent if variational method~\cite{Tinkham_book} or other means are used to derive the ground state.   Refs.~\cite{Unruh_book,Blasone_book} show a generalized entropy can be defined and evaluated if the Bogoliubov transformation is implemented for constructing thermal vacuum of fermions. Here we show that, by analyzing the Bogoliubov transformation for the BCS ground state, the generalized entropy is the entanglement entropy of the BCS ground state, and its explicit expression will be presented. Recent experiments measuring the Renyi entropy \cite{Renyi61,RenyiS}, which also reveals entanglement inside quantum systems, may provide hints for future probes of the entanglement entropy of many-body systems.

The Berry phase \cite{Berry84} measures how the geometrical phase of the wave function changes when a system is adiabatically transformed along a loop in the parameter space. By tuning the phase of the order parameter continuously, the Berry phase of the generalized BCS ground state can be found. We will show that it is $N\pi$ when the phase of the order parameter is changed by $2\pi$, where $N$ is the averaged number of fermions in the system. Refs.~\cite{Volovik_book,Blasone_book} show that the Berry phase can be related to the magnetic flux of effective monopoles in the corresponding parameter space, and we will follow this relation to find the effective monopole associated with the BCS ground state. Interestingly, the Dirac quantization condition connecting the electric and magnetic charges is satisfied. Measurements of the Zak phase \cite{Zakphase}, which is the Berry phase of one dimensional systems, of cold-atoms in optical superlattices have been performed and shed light on quantum phases in topological systems.

Moreover, the generalized BCS ground state and its Bogoliubov transformation will be shown to follow a  $q$-deformed Hopf superalgebra, which is similar to an algebra applied to noninteracting fermions \cite{VPLA98,Blasone_book}. The ground state of noninteracting fermions is a Fermi sea, and Refs.~\cite{VPLA98,Unruh_book,Blasone_book} use a Bogoliubov transformation to construct the corresponding thermal vacuum for demonstrating inequivalent representations. The algebraic structures will also help clarify the origin of similarities between different superfluid ground states.

In a similar manner, the ground state of noninteracting bosons can be expressed as an SU(1,1) generalized coherent state~\cite{Perelomov_book}. Here we generalize the algebraic structure and apply it to a weakly-interacting Bose gas. The ground state corresponds to a condensate of Bose-Einstein condensation (BEC), and we successfully obtained the explicit expression by performing a Bogoliubov transformation of bosons.
Due to the Bogoliubov transformations, both the BCS and BEC ground states exhibit the feature of two copies of degrees of freedom naturally embedded in their structures. Such a bipartite feature has been discussed in inequivalent representations of quantum field theory or thermal field dynamics \cite{VPLA98,Unruh_book,Blasone_book}, but in those works the two copies have to be introduced.

The generalized uncertainty relation, Berry phase and its associated monopole, entanglement entropy, and algebraic structure of the BEC ground state are also analyzed, and the results are similar to the BCS case with subtle differences. For example, the Berry phase of the BEC ground state is proportional to only the excited bosons outside the condensate, in contrast to the BCS case where the Berry phase is proportional to the total particle number. The Hamiltonians of the fermionic BCS superfluids and the weakly-interacting Bose gases both have a family of generalized ground states, and the Bogoliubov transformations generalizing them may be classified by the $q$-deformed Hopf algebra.

The paper is organized as follows. Sec.~\ref{sec:BCS} presents the fermionic BCS ground state and its algebraic structure. The Bogoliubov transformation for deriving the ground state, generalized uncertainty relation, Berry phase, entanglement entropy, and $q$-deformed Hopf algebra will be discussed. We also compare the BCS wave function with the fixed particle-number wave function and contrast their similarities and differences. Sec.~\ref{sec:BEC} presents a similar discussion on the bosonic BEC
ground state. Sec.~\ref{sec:conclusion} concludes our work. Some details of the calculations and the fixed particle-number wavefunction of fermionic superfluids are summarized in the Appendix.

\section{BCS superfluids}\label{sec:BCS}
We first discuss the BCS ground state and its algebraic properties, and a parallel discussion on weakly interacting Bose gases will be presented next.

\subsection{Fermionic BCS ground state}
After applying the BCS mean-field approximation \cite{Walecka}, the second-quantized Hamiltonian takes the form
\begin{eqnarray}\label{HBCS}
H_{\textrm{BCS}}&=&\sum_{\mathbf{k}\sigma}\psi^{\dagger}_{\mathbf{k}\sigma}\Big(\frac{\mathbf{k}^2}{2m}-\mu\Big)\psi_{\mathbf{k}\sigma}
+\sum_{\mathbf{k}}\Delta^*\psi_{-\mathbf{k}\uparrow}\psi_{\mathbf{k}\downarrow} \nonumber \\
& &+\sum_{\mathbf{k}}\Delta\psi^{\dagger}_{\mathbf{k}\downarrow}\psi^{\dagger}_{-\mathbf{k}\uparrow}+\frac{|\Delta|^2}{g},
\end{eqnarray}
where $\psi^{\dagger}$ and $\psi$ are the fermion creation and annihilation operators, $\sigma=\uparrow,\downarrow$ labels the spin, $\mu$ is the chemical potential, and we set $\hbar=1$.   $\Delta(\mathbf{x})=g\langle\psi_{\uparrow}(\mathbf{x})\psi_{\downarrow}(\mathbf{x})\rangle$ is the gap function, which is equivalent to the order parameter, and $g$ is the coupling constant. $\Delta$ is constant if the system is uniform and static. The ground state can be constructed via a variational method~\cite{BCS57,Tinkham_book}
\begin{eqnarray}\label{GS0}
|\textrm{BCS} \rangle
&=&\prod_{\mathbf{k}}\big(u_{\mathbf{k}}+v^*_{\mathbf{k}}\psi^{\dagger}_{\mathbf{k}\uparrow}\psi^{\dagger}_{-\mathbf{k}\downarrow}\big)|0\rangle,
\end{eqnarray}
where $u_{\mathbf{k}},v_{\mathbf{k}}$ are chosen to minimize the ground state energy. Here we will generalize the Bogoliubov transformation equivalent to a canonical transformation \cite{Walecka} to elucidate the algebraic structure of the BCS ground state.

Although one can use the U(1) symmetry to choose $\Delta$ to be real, here we consider the general case where $\Delta$ is  complex and will later show that its phase indeed brings nontrivial physical effects. The BCS Hamiltonian (\ref{HBCS}) can be diagonalized by the Bogoliubov transformation \cite{Walecka}
\begin{eqnarray}\label{BT1}
\alpha_{\mathbf{k}}=u^{\ast}_{\mathbf{k}}\psi_{\mathbf{k}\uparrow}-v^{\ast}_{\mathbf{k}}\psi^{\dagger}_{-\mathbf{k}\downarrow},\quad \beta^{\dagger}_{-\mathbf{k}}=v_{\mathbf{k}}\psi_{\mathbf{k}\uparrow}+u_{\mathbf{k}}\psi^{\dagger}_{-\mathbf{k}\downarrow}.
\end{eqnarray}
The linear transformation is canonical if and only if the new operators obey the relations
\begin{eqnarray}\label{AQ1}
\{\alpha_{\mathbf{k}},\alpha^{\dagger}_{\mathbf{k}'}\}=\{\beta_{\mathbf{k}},\beta^{\dagger}_{\mathbf{k}'}\}=\delta_{\mathbf{k}\mathbf{k}'},
\end{eqnarray}
and all other anticommutations vanish.
Given the anticommutation relations
$\{\psi_{\mathbf{k}\sigma},\psi^{\dagger}_{\mathbf{k}'\sigma'}\}=\delta_{\mathbf{k}\mathbf{k}'}\delta_{\sigma\sigma'}$, Eq.~\eqref{AQ1} implies $|u_{\mathbf{k}}|^2+|v_{\mathbf{k}}|^2=1$.
To diagonalize the Hamiltonian, the coefficients $u_{\mathbf{k}}$ and $v_{\mathbf{k}}$ can be chosen to satisfy
\begin{eqnarray}\label{BT20}
|u_{\mathbf{k}}|^2-|v_{\mathbf{k}}|^2=\frac{\xi_{\mathbf{k}}}{E_{\mathbf{k}}},\quad
u_{\mathbf{k}}v^*_{\mathbf{k}}=\frac{\Delta}{2E_{\mathbf{k}}},
\end{eqnarray}
where $\xi_{\mathbf{k}}=\frac{\mathbf{k}^2}{2m}-\mu$, $E_{\mathbf{k}}=\sqrt{\xi^2_{\mathbf{k}}+|\Delta|^2}$ is the excitation energy of the quasi-particles.
To fully determine $u_{\mathbf{k}}$ and $v_{\mathbf{k}}$, we need four equations. However, these two equations together with the canonical condition are enough to diagonalize the Hamiltonian, and
the BCS Hamiltonian becomes
\begin{equation}\label{HBCS2}
H_{\textrm{BCS}}=\sum_{\mathbf{k}}E_{\mathbf{k}}(\alpha^{\dagger}_{\mathbf{k}}\alpha_{\mathbf{k}}+\beta^{\dagger}_{-\mathbf{k}}\beta_{-\mathbf{k}})+\sum_{\mathbf{k}}(\xi_{\mathbf{k}}-E_{\mathbf{k}})+\frac{|\Delta|^2}{g}.
\end{equation}
Thus, $\alpha_{\mathbf{k}}$ and $\beta_{\mathbf{k}}$ are the annihilation operators for the quasi-particles, which mix particles and antiparticles (holes). The total particle number $N$ and the gap $\Delta$ are determined self-consistently by  $N=\sum_{\mathbf{k},\sigma}\langle\psi^{\dagger}_{\mathbf{k}\sigma}\psi_{\mathbf{k}\sigma}\rangle$ and $\Delta=g\sum_{\mathbf{k}}\langle\psi_{\mathbf{k}\uparrow}\psi_{-\mathbf{k},\downarrow} \rangle$. With the diagonalized Hamiltonian, one obtains
\begin{equation}\label{geBCS}
N=\sum_{\mathbf{k}}\Big[1-\frac{\xi_{\mathbf{k}}}{E_{\mathbf{k}}}\big(1-2f(E_{\mathbf{k}})\big)\Big],~ -\frac{1}{g}=\sum_{\mathbf{k}}\frac{1-2f(E_{\mathbf{k}})}{2E_{\mathbf{k}}}.
\end{equation}
Here $f(x)$ is the Fermi-Dirac distribution function. At zero temperature they become $N=2\sum_{\mathbf{k}}|v_{\mathbf{k}}|^2$ and $-\frac{1}{g}=\sum_{\mathbf{k}}\frac{1}{2E_{\mathbf{k}}}$.

The Bogoliubov transformation can be rewritten with the parametric representation $u_{\mathbf{k}}=\cos\theta_{\mathbf{k}}$, $v_{\mathbf{k}}=\sin\theta_{\mathbf{k}}e^{i\tau}$.
The phase $\tau$ is introduced since we are left with one degree of freedom for solving $u_{\mathbf{k}}$ and $v_{\mathbf{k}}$.
Eq.(\ref{BT20}) indicates
\begin{eqnarray} \label{B0}
\sin2\theta_{\mathbf{k}}=\frac{\Delta}{E_{\mathbf{k}}}e^{i\tau},\qquad \cos2\theta_{\mathbf{k}}=\frac{\xi_{\mathbf{k}}}{E_{\mathbf{k}}},
\end{eqnarray}
then it leads to $\Delta=|\Delta|e^{-i\tau}$. Thus, $-\tau$ is the phase of the order parameter. The Bogoliubov transformation (\ref{BT1}) now takes the form
\begin{eqnarray}\label{BT2}
\alpha_{\mathbf{k}}&=&e^G\psi_{\mathbf{k}\uparrow}e^{-G},~\beta^{\dagger}_{-\mathbf{k}}=e^G\psi^{\dagger}_{-\mathbf{k}\downarrow}e^{-G}.
\end{eqnarray}
Here the generator is $G=\sum_{\mathbf{k}}\theta_{\mathbf{k}}\big(\psi_{\mathbf{k}\uparrow}\psi_{-\mathbf{k}\downarrow}e^{i\tau}+\psi^{\dagger}_{\mathbf{k}\uparrow}\psi^{\dagger}_{-\mathbf{k}\downarrow}e^{-i\tau}\big)$.
The transformation $e^G$ is unitary because $G^{\dagger}=-G$.
Since the Bogoliubov transformation is linear between $(\psi_{\mathbf{k}\uparrow}, \psi_{-\mathbf{k}\downarrow})$ and $(\alpha_{\mathbf{k}},\beta_{-\mathbf{k}})$, the Fock-space vacuum (i.e., ground state) of the $\psi_{\mathbf{k}\sigma}$-quantum is transformed into the Fock-space vacuum of the $\alpha$ and $\beta$-quantum. We define the former as $|0\rangle$ and the latter as $|0(\theta)\rangle\equiv e^{G}|0\rangle$, then we have $\alpha_{\mathbf{k}}|0(\theta)\rangle=e^G\psi_{\mathbf{k}\uparrow}|0\rangle=0$, $\beta_{-\mathbf{k}}|0(\theta)\rangle=e^G\psi_{-\mathbf{k}\downarrow}|0\rangle=0$.
Since the energy dispersions of the quasi-particles $\alpha$- and $\beta$-quanta are positive, the Fock-space vacuum of the $\alpha$- and $\beta$-quanta, $|0(\theta)\rangle$, is also the ground state of the BCS Hamiltonian.
We follow similar methods of Ref.~\cite{Ezawa_book} to derive the generalized form of BCS ground state.
The relation $\alpha_{\mathbf{k}}|0(\theta)\rangle=\beta_{-\mathbf{k}}|0(\theta)\rangle=0$ leads to
$\big(\frac{\partial}{\partial\psi^{\dagger}_{\mathbf{k}\uparrow}}-\tan\theta_{\mathbf{k}}e^{-i\tau}\psi^{\dagger}_{-\mathbf{k}\downarrow}\big)|0(\theta)\rangle=0$ and  $\big(\frac{\partial}{\partial\psi^{\dagger}_{-\mathbf{k}\downarrow}}+\tan\theta_{\mathbf{k}}e^{-i\tau}\psi^{\dagger}_{\mathbf{k}\uparrow}\big)|0(\theta)\rangle=0$.
The solution is
\begin{eqnarray}\label{GS1}
|0(\theta)\rangle&=&C\prod_{\mathbf{k}}e^{\tan\theta_{\mathbf{k}}e^{-i\tau}\psi^{\dagger}_{\mathbf{k}\uparrow}\psi^{\dagger}_{-\mathbf{k}\downarrow}}|0\rangle.
\end{eqnarray}
Here $C=C'\prod_{\mathbf{k}}u_{\mathbf{k}}$ and $C'=1$ as shown in Appendix \ref{appBCS}.
The above derivation avoids a possible ambiguity about the signs of $u_{\mathbf{k}}$, and the result agrees with the original BCS ground state \eqref{GS0} when $\tau$ is set to $0$.

The BCS ground state is a generalized coherent state. To see this we introduce the operators $S^+_{\mathbf{k}}=\psi^{\dagger}_{-\mathbf{k}\downarrow}\psi^{\dagger}_{\mathbf{k}\uparrow}$, $S^-_{\mathbf{k}}=\psi_{\mathbf{k}\uparrow}\psi_{-\mathbf{k}\downarrow}$, and $S^z_{\mathbf{k}}=\frac{1}{2}(\psi^{\dagger}_{\mathbf{k}\uparrow}\psi_{\mathbf{k}\uparrow}+\psi^{\dagger}_{-\mathbf{k}\downarrow}\psi_{-\mathbf{k}\downarrow}-1)$.
It can be shown that $[S^+_{\mathbf{k}},S^-_{\mathbf{k}}]=2S^z_{\mathbf{k}}$ and $[S^z_{\mathbf{k}},S^{\pm}_{\mathbf{k}}]=\pm S^{\pm}_{\mathbf{k}}$, so they form an SU(2) algebra. However, $(S^+_{\mathbf{k}})^2=(S^-_{\mathbf{k}})^2=0$, so it is actually a second-order nilpotent SU(2) algebra. The state transformation $|0(\theta)\rangle =e^{G}|0\rangle$ can be expressed as
\begin{eqnarray}\label{GS3}
|0(\theta)\rangle&=&e^{\sum_{\mathbf{k}}(\tau_{\mathbf{k}}S^+_{\mathbf{k}}-\tau^*_{\mathbf{k}}S^-_{\mathbf{k}})}|0\rangle \nonumber \\
&=&e^{-\frac{1}{2}\sum_{\mathbf{k}}\ln(1+|\zeta_{\mathbf{k}}|^2)}e^{\sum_{\mathbf{k}}\zeta_{\mathbf{k}}S^+_{\mathbf{k}}}|0\rangle.
\end{eqnarray}
Here $\tau_{\mathbf{k}}=-\theta_{\mathbf{k}}e^{-i\tau}$ and $\zeta_{\mathbf{k}}=\zeta_{\mathbf{k}}(\tau_{\mathbf{k}})\equiv\frac{\tau_{\mathbf{k}}\tan(|\tau_{\mathbf{k}}|)}{|\tau_{\mathbf{k}}|}$.
This defines a second-order nilpotent SU(2) generalized coherent state~\cite{Perelomov_book}. In obtaining the expression, we have used the Baker-Campbell-Hausdorff disentangling formula \cite{Campbell1897,Poincare1899,Baker02} and $S^z_{\mathbf{k}}|0\rangle=-\frac{1}{2}|0\rangle$, which can also be obtained directly from Eq.~(\ref{GS1}).

The generalized BCS ground state is also a generalized squeezed state. We introduce the pseudo angular momentum operators $S^x_{\mathbf{k}}=\frac{1}{2}(S^+_{\mathbf{k}}+S^-_{\mathbf{k}})$, $S^y_{\mathbf{k}}=\frac{1}{2i}(S^+_{\mathbf{k}}-S^-_{\mathbf{k}})$
which satisfy the relation $[S^x_{\mathbf{k}},S^y_{\mathbf{k}}]=iS^z_{\mathbf{k}}$.
One can show that the BCS ground state saturates the Robertson-Schrodinger inequality ~\cite{Merzbacher_book} with an equal sign for any $\mathbf{k}$ (see Appendix~\ref{appBCS} for details):
\begin{eqnarray}\label{eq:RSine}
\textrm{cov}(S^x_{\mathbf{k}},S^x_{\mathbf{k}})\textrm{cov}(S^y_{\mathbf{k}},S^y_{\mathbf{k}})&\ge& \frac{1}{4}|\langle0(\theta)|[S^x_{\mathbf{k}},S^y_{\mathbf{k}}]|0(\theta)\rangle|^2 \nonumber \\
& &+ \textrm{cov}^2
(S^x_{\mathbf{k}},S^y_{\mathbf{k}}),
\end{eqnarray}
where
$\textrm{cov}(A,B)=\frac{1}{2}\langle \psi |(AB+BA)|\psi\rangle-\langle \psi |A|\psi\rangle\langle \psi |B|\psi\rangle$ for any state $|\psi\rangle$ and operators $A, B$.
The above inequality reduces to the well-known Heisenberg uncertainty relation when the last term is absent.

\subsection{Berry phase of BCS ground state}
By examining Eqs.~(\ref{HBCS2}) and (\ref{B0}), one can see that while the BCS Hamiltonian only depends on $|\Delta|$, the ground state depends on the amplitude and phase of $\Delta$. Hence, there is a family of ground states which forms a U(1) manifold. When a fixed phase of the condensate is chosen, the U(1) symmetry of the BCS Hamiltonian is spontaneously broken. Here we consider the order parameter can be perturbed externally. If the phase is perturbed while the amplitude is intact, the system is transported along the U(1) manifold. We will take the complex $\Delta$ plane as the parameter space of the BCS theory, so a transportation along the U(1) manifold is equivalent to varying $\Delta$ along a unit circle on the complex $\Delta$ plane by external fields.

A quantized Berry phase of the BCS ground state emerges when the order parameter is varied along a closed path in the parameter space. The Berry phase for the SU(1,1) group has been discussed in Ref.~\cite{FHY88} and we will discuss its relation to the BEC ground state later, but we have not found an expression of the SU(2) group. Hence, we will present a derivation of the Berry phase for the SU(2) generalized coherent state, and this applies to the BCS ground state because it is a second-order nilpotent SU(2) coherent state.

We parameterize a circle $C$ in the parameter space by $\tau$, the phase of the order parameter, with $0\le\tau < 2\pi$. $\theta_{\mathbf{k}}$ is kept a constant along the trajectory for any $\mathbf{k}$ since $|\Delta|$ is fixed. The Berry phase is given by~\cite{Berry84}
\begin{eqnarray}
\gamma(C)=i\int_0^{2\pi}d\tau\langle 0(\theta(\tau))|\frac{d}{d\tau}|0(\theta(\tau))\rangle,
\end{eqnarray}
where $|0(\theta(\tau))\rangle$ and $\zeta_{\mathbf{k}}$ are shown in Eq.~(\ref{GS3}) and its discussion.
The Berry phase is then evaluated as
\begin{eqnarray}
\gamma(C)&=&i\int_0^{2\pi} d\tau \sum_{\mathbf{k}}\left(-\frac{1}{2}\frac{\zeta^*_{\mathbf{k}}\frac{d\zeta_{\mathbf{k}}}{d\tau}+\zeta_{\mathbf{k}}\frac{d\zeta^*_{\mathbf{k}}}{d\tau}}{1+|\zeta_{\mathbf{k}}|^2} \right. \nonumber \\
& &\left. +\langle 0(\theta(\tau))|S^+_{\mathbf{k}}|0(\theta(\tau))\rangle\frac{d\zeta_{\mathbf{k}}}{d\tau}\right)\nonumber\\
&=&\frac{i}{2}\sum_{\mathbf{k}}\oint_C\frac{\zeta^*_{\mathbf{k}}d\zeta_{\mathbf{k}}-\zeta_{\mathbf{k}}d\zeta^*_{\mathbf{k}}}{1+|\zeta_{\mathbf{k}}|^2}\nonumber\\
&=&\sum_{\mathbf{k}}\int_0^{2\pi}d\tau\sin^2\theta_{\mathbf{k}}(\tau)\nonumber\\
&=&\sum_{\mathbf{k}}\int_0^{2\pi}d\tau|v_{\mathbf{k}}(\tau)|^2.
\end{eqnarray}
In the second line we have used the identity  $\langle0(\theta)|S^+_{\mathbf{k}}|0(\theta)\rangle=\frac{\zeta^*_{\mathbf{k}}}{1+|\zeta_{\mathbf{k}}|^2}$.
At zero temperature, the total fermion number is $N=2\sum_{\mathbf{k}}|v_{\mathbf{k}}|^2$, so
\begin{eqnarray}\label{BPBCS}
\gamma(C)=\frac{1}{2}N\int_0^{2\pi}d\tau=N\pi,
\end{eqnarray}
Interestingly, the Berry phase is proportional to the total particle number and is an integer multiple of $\pi$.

The Berry phase can also be obtained by constructing an effective vector potential in the $\Delta-\tau$ parameter space, where $\Delta$ may be perturbed by the chemical potential $\mu$. One can directly verify that the Berry phase satisfies
\begin{equation}
\gamma(C)=\sum_{\mathbf{k}}\int_0^{2\pi}d\tau\sin^2\theta_{\mathbf{k}}=\frac{1}{2}\sum_{\mathbf{k}}\int_0^{2\pi}d\tau\frac{x_{\mathbf{k}}dy_{\mathbf{k}}-y_{\mathbf{k}}dx_{\mathbf{k}}}{E_{\mathbf{k}}(z_{\mathbf{k}}+E_{\mathbf{k}})},
\end{equation}
where (see Eqs.(\ref{B0}))
\begin{eqnarray}
& &x_{\mathbf{k}}=E_{\mathbf{k}}\sin(2\theta_{\mathbf{k}})\cos\tau=\textrm{Re}\Delta,\nonumber\\
& &y_{\mathbf{k}}=E_{\mathbf{k}}\sin(2\theta_{\mathbf{k}})\sin\tau=-\textrm{Im}\Delta,\nonumber\\
& &z_{\mathbf{k}}=E_{\mathbf{k}}\cos(2\theta_{\mathbf{k}})=\xi_{\mathbf{k}}.
\end{eqnarray}
Therefore, we can define the vector potential, or the Berry's connection, as
$\mathbf{A}_{\mathbf{k}}=\frac{1}{2}\frac{(-y_{\mathbf{k}},x_{\mathbf{k}},0)}{E_{\mathbf{k}}(z_{\mathbf{k}}+E_{\mathbf{k}})}
=\frac{1}{2}\frac{(\textrm{Im}\Delta,\textrm{Re}\Delta,0)}{E_{\mathbf{k}}(\xi_{\mathbf{k}}+E_{\mathbf{k}})}$. In  spherical coordinates it becomes
\begin{eqnarray}
\mathbf{A}_{\mathbf{k}}=\frac{1}{2}(0,0,\frac{1-\cos2\theta_{\mathbf{k}}}{E_{\mathbf{k}}\sin2\theta_{\mathbf{k}}})
=\frac{1}{2}(0,0,\frac{E_{\mathbf{k}}-\xi_{\mathbf{k}}}{E_{\mathbf{k}}|\Delta|}).
\end{eqnarray}
Here the radial distance is $E_{\mathbf{k}}$, the polar angle is $2\theta_{\mathbf{k}}$. and the azimuthal angle is $\tau$.
The Berry phase is then the line integral $\sum_{\mathbf{k}}\int_0^{2\pi}d\tau \mathbf{A}_{\mathbf{k}}\cdot d\mathbf{x}_{\mathbf{k}}$. We remark that the vector potential has no singularity in the parameter space as long as $|\Delta|>0$.

The corresponding Berry curvature, which is an effective magnetic field in the parameter space, is
\begin{eqnarray}
\mathbf{B}_{\mathbf{k}}=\nabla\times\mathbf{A}_{\mathbf{k}}=\frac{1}{2}\frac{\mathbf{x}_{\mathbf{k}}}{E^3_{\mathbf{k}}}=\frac{1}{2}\frac{\mathbf{n}_{\mathbf{k}}}{E^2_{\mathbf{k}}},
\end{eqnarray}
where $\mathbf{n}_{\mathbf{k}}=(\sin2\theta_{\mathbf{k}}\cos\tau,\sin2\theta_{\mathbf{k}}\sin\tau,\cos2\theta_{\mathbf{k}})$. This is the magnetic field from a monopole with a magnetic charge $\frac{1}{2}$ (in natural units) located at the origin of the parametric space. Pictorially, one may follow Ref.~\cite{Volovik_book} and references therein to think of the Berry phase as the flux carried by a Dirac string attached the monopole.
Therefore, the Berry phase is also equal to the magnetic flux of the monopole:
\begin{eqnarray} \label{BCP}
\gamma(C)=\sum_{\mathbf{k}}\iint_S\mathbf{B}_{\mathbf{k}}\cdot d\mathbf{S}_{\mathbf{k}}=N\pi,
\end{eqnarray}
The Dirac quantization condition $Q_e Q_m=1$ in natural units guarantees the single-valueness of wave functions around the Dirac string~\cite{Volovik_book}, where $Q_e$ and $Q_m$ are the elementary electric and magnetic charges. In BCS theory, a Cooper pair carries electric charge $Q^{\textrm{BCS}}_{\textrm{e}}=2$ due to the two constituent fermions. However, the effective magnetic charge of a monopole from the BCS ground state is only $Q^{\textrm{BCS}}_{\textrm{m}}=\frac{1}{2}$. Thus, the Dirac quantization condition is respected and the Dirac string is invisible to the fermions \cite{Volovik_book}.

\subsection{Entanglement entropy of BCS ground state}
The BCS ground state can not be factorized into a product of single-mode states, hence it possesses entanglement between the $\psi_{\sigma}$-quanta.
To measure the entanglement of a bipartite system with components A and B, a generalization of the von Neumann entropy of a subsystem is introduced~\cite{QP_book}. An entangled state has a Schmidt decomposition
\begin{eqnarray}
|\Psi\rangle=\sum_ic_i|\psi_i\rangle_{\textrm{A}}\otimes|\phi_i\rangle_{\textrm{B}},
\end{eqnarray}
with $c_i\ge0$ and $\sum_ic^2_i=1$. The associated density matrix is $ \rho=|\Psi\rangle\langle\Psi|.$ By taking partial trace over the basis of one subsystem, the reduced density matrix is obtained. For example, $\rho_{\textrm{A}}=\textrm{Tr}_{\textrm{B}}(\rho)$.
The von Neumann entropy of subsystem A gives the entanglement entropy \cite{Horodecki09} $ S=-\textrm{Tr}_{\textrm{A}}(\rho_{\textrm{A}}\ln\rho_{\textrm{A}})$.
A straightforward evaluation shows
\begin{eqnarray}\label{E1}
S=-\sum_ic^2_i\ln c^2_i.
\end{eqnarray}
A pure state of the composite system may lead to effective mixed states of its subsystems, but the entanglement entropies of the two subsystems are the same \cite{Susskind_book}.

From Eq.~(\ref{GS1}), the BCS ground state can be expressed as a Schmidt decomposition (with details shown in Appendix~\ref{appBCS})
\begin{eqnarray}\label{GS2}
	|0(\theta)\rangle&=&\prod_{\mathbf{k}}\cos\theta_{\mathbf{k}}e^{\tan\theta_{\mathbf{k}}e^{-i\tau}\psi^{\dagger}_{\mathbf{k}\uparrow}\psi^{\dagger}_{-\mathbf{k}\downarrow}}|0\rangle  \\
	&=&\sum_{n=0}^\infty e^{-in\tau}\sum_{\mathbf{k};\{n\}}\sqrt{W_{\mathbf{k};\{n\}}}\prod_{\mathbf{k};\{n\}'}|n_{\mathbf{k}\uparrow}\rangle\otimes|n_{-\mathbf{k}\downarrow}\rangle, \nonumber
\end{eqnarray}
where the occupation number $n_{\mathbf{k}}=0$ or $1$, $|n_{\mathbf{k}\sigma}\rangle=(\psi^{\dagger}_{\mathbf{k}\sigma})^{n_{\mathbf{k}}}|0\rangle$, and the notation $\sum_{\mathbf{k};\{n\}}$ means the summation is taken over all possible $\mathbf{k}$ only when $n_\mathbf{k}=1$ and  $\sum_{\mathbf{k}}n_\mathbf{k}=n$. Here the amplitude $W_{\mathbf{k};\{n\}}=\prod_{\mathbf{k};\{n\}'}\sin^{2n_\mathbf{k}}\theta_\mathbf{k}\cos^{2(1-n_\mathbf{k})}\theta_\mathbf{k}$,
where $\prod_{\mathbf{k};\{n\}'}$ means the multiplication is taken over all $\mathbf{k}$ such that $\sum_{\mathbf{k}}n_{\mathbf{k}}=n$ with $n_{\mathbf{k}}=0$ or $1$. Since $0\le W_{\mathbf{k};\{n\}}\le 1$, $W_{\mathbf{k};\{n\}}$ is non-negative. For simplicity, we use the sub index $\mathbf{k};\{n\}$ to label the summations or products throughout the rest of the paper.
It can be shown that
\begin{eqnarray}
	\sum_{n=0}^{\infty}\sum_{\mathbf{k};\{n\}}W_{\mathbf{k};\{n\}}
	&=&\prod_{\mathbf{k}}\sum_{n_{\mathbf{k}}=0}^1\sin^{2n_{\mathbf{k}}}\theta_{\mathbf{k}}\cos^{2(1-n_{\mathbf{k}})}\theta_{\mathbf{k}}\nonumber\\
	&=&1.
\end{eqnarray}
Thus, $W_{\mathbf{k};\{n\}}$ is the probability of  $\prod_{\mathbf{k};\{n\}'}(|n_{\mathbf{k}\uparrow}\rangle\otimes|n_{-\mathbf{k}\downarrow}\rangle)$ in the BCS ground state. From this decomposition, the BCS ground state is shown to possess entanglement not only between
the $\psi_{\sigma}$ quanta, but also between different momentum distribution of a $n$-pair state.

Since the ground state \eqref{GS2} has the structure of a Schmidt decomposition, we can take $\rho=|0(\theta)\rangle\langle0(\theta)|$ and $\textrm{A}=\uparrow$, $\textrm{B}=\downarrow$. Hence,
\begin{eqnarray}
\rho_{\uparrow}=\sum_{n=0}^{\infty}\sum_{\mathbf{k};\{n\}}W_{\mathbf{k};\{n\}}\prod_{\mathbf{k};\{n\}'}|n_{\mathbf{k}\uparrow}\rangle\langle n_{\mathbf{k}\uparrow}|.
\end{eqnarray}
Using Eq.(\ref{E1}), the entanglement entropy between the two spin states is then expressed as  $S=-\textrm{Tr}_\uparrow(\rho_\uparrow\ln\rho_\uparrow)=-\sum_{n=0}^\infty\sum_{\mathbf{k};\{n\}}W_{\mathbf{k};\{n\}}\ln W_{\mathbf{k};\{n\}}$. By plugging in the expression of $W_{\mathbf{k};\{n\}}$ one can get the entanglement entropy. However, a simpler expression of it can be obtained by using the relation $S=-\textrm{Tr}_\uparrow(\rho_\uparrow\ln\rho_\uparrow)=-\lim_{n\rightarrow1^+}\frac{d}{dn}\textrm{Tr}_\uparrow\rho_\uparrow^n=-\lim_{n\rightarrow1^+}\frac{d}{dn}\sum_{m=0}^{\infty}\sum_{\mathbf{k};\{m\}}W^n_{\mathbf{k};\{m\}}$. Explicitly, we have
\begin{eqnarray}\label{EBCS2}
S
&=&-\lim_{n\rightarrow1^+}\frac{d}{dn}\sum_{m=0}^{\infty}\sum_{\mathbf{k};\{m\}}\prod_{\mathbf{k};\{m\}'}\sin^{2m_{\mathbf{k}}n}\theta_{\mathbf{k}}\cos^{2n(1-m_{\mathbf{k}})}\theta_{\mathbf{k}}\nonumber\\
&=&-\lim_{n\rightarrow1^+}\frac{d}{dn}\prod_{\mathbf{k}}\sum_{m_{\mathbf{k}}=0}^1\sin^{2nm_{\mathbf{k}}}\theta_{\mathbf{k}}\cos^{2n(1-m_{\mathbf{k}})}\theta_{\mathbf{k}}\nonumber\\
&=&-\sum_{\mathbf{k}}(|u_{\mathbf{k}}|^{2}\ln|u_{\mathbf{k}}|^{2}+|v_{\mathbf{k}}|^{2}\ln|v_{\mathbf{k}}|^{2})\nonumber\\
&=&-\sum_{\mathbf{k}}\Big(\frac{E_{\mathbf{k}}+\xi_{\mathbf{k}}}{2E_{\mathbf{k}}}\ln\frac{E_{\mathbf{k}}+\xi_{\mathbf{k}}}{2E_{\mathbf{k}}}+
\frac{E_{\mathbf{k}}-\xi_{\mathbf{k}}}{2E_{\mathbf{k}}}\ln\frac{E_{\mathbf{k}}-\xi_{\mathbf{k}}}{2E_{\mathbf{k}}}\Big).\nonumber\\
\end{eqnarray}
This entropy quantifies the internal entanglement of a many-body ground state \cite{Unruh_book}. Since $0<\frac{E_{\mathbf{k}}\pm\xi_{\mathbf{k}}}{2E_{\mathbf{k}}}<1$, the entanglement entropy is finite for the BCS ground state.

Interestingly, the entanglement entropy can also be obtained by taking the vacuum expectation value of the entropy operator either for the spin-up component given by
\begin{eqnarray}\label{Su1}
	S_{\uparrow}
 &=&-\sum_{\mathbf{k}}(\psi^{\dagger}_{\mathbf{k}\uparrow}\psi_{\mathbf{k}\uparrow}\ln\tan^2\theta_{\mathbf{k}}+\ln\cos^2\theta_{\mathbf{k}})
\end{eqnarray}
or for the spin-down component defined in a similar way. Similar expressions have been  presented in Ref.~\cite{Unruh_book} for non-superfluid states, where the concept of doubling the degrees of freedom and Eq.~(\ref{E1}) are implemented. In contrast, the BCS superfluid naturally has two sets of entangled operators, so there is no need for doubling its degrees of freedom. Thus, the entanglement entropy is intrinsic in the wavefunction.

\subsection{Algebraic structure of BCS ground state}
If the amplitude of the order parameter is perturbed while the phase is kept unchanged, it can be shown that the Bogoliubov transformation which generates the BCS ground state can be described by the $q$-deformation of the $Z_2$-graded Hopf algebra (which is actually a Hopf superalgebra) ~\cite{VPLA98}  $\bigoplus_{\mathbf{k}}h_{q_{\mathbf{k}}}(1|1)$, where $q_{\mathbf{k}}=e^{2i\theta_{\mathbf{k}}}$. Experimentally, the pairing gap can be tuned in a two-component ultra-cold atomic Fermi gas, where the inter-particle interaction can be adjusted by a Feshbach resonance \cite{Feshbach58}. We give a brief discussion on the algebraic properties here.

The BCS ground state is $|0(\theta)\rangle=e^{\sum_{\mathbf{k}}\theta_{\mathbf{k}}\mathcal{G}_{\mathbf{k}}}
|0\rangle$ with $\mathcal{G}_{\mathbf{k}}=\psi_{\mathbf{k}\uparrow}\psi_{-\mathbf{k}\downarrow}e^{i\tau}+\psi^{\dagger}_{\mathbf{k}\uparrow}\psi^{\dagger}_{-\mathbf{k}\downarrow}e^{-i\tau}$ following Eq.~\eqref{BT2}. If only the amplitude of the order parameter is perturbed so that $\theta_{\mathbf{k}}\rightarrow\theta'_{\mathbf{k}}$, then the new ground state becomes $|0(\theta')\rangle\equiv e^{\sum_{\mathbf{k}}\theta'_{\mathbf{k}}\mathcal{G}_{\mathbf{k}}}|0\rangle$. Hence,
$\langle 0(\theta')|0(\theta)\rangle=\langle 0|e^{\sum_{\mathbf{k}}(\theta_{\mathbf{k}}-\theta_{\mathbf{k}}')\mathcal{G}_{\mathbf{k}}}|0\rangle=e^{\frac{V}{(2\pi)^3}\int d^3\mathbf{k}\ln|\cos(\theta_{\mathbf{k}}-\theta'_{\mathbf{k}})|}$.
Since $|\cos(\theta_{\mathbf{k}}-\theta'_{\mathbf{k}})|\le 1$, $\langle 0(\theta')|0(\theta)\rangle=0$ in the thermodynamic limit when $V\rightarrow\infty$. This is because different values of $\theta$ denote inequivalent representations of the BCS ground states but there is no tunneling between them in the thermodynamic limit. Moreover,
\begin{eqnarray}
e^{\sum_{\mathbf{k}}\theta'_{\mathbf{k}}\mathcal{G}_{\mathbf{k}}}|0(\theta)\rangle=e^{\sum_{\mathbf{k}}(\theta_{\mathbf{k}}+\theta'_{\mathbf{k}})\mathcal{G}_{\mathbf{k}}}|0\rangle=|0(\theta+\theta')\rangle.
\end{eqnarray}
implies a theta-vacuum~\cite{Srednicki_book} like structure, and the Bogoliubov transformation generates translations among those inequivalent BCS ground state representations. It also preserves the generalized minimum uncertainty relation \eqref{eq:RSine}.

Inspired by the ideas of duplicating the degrees of freedom in Refs.~\cite{VPLA98,Unruh_book}, we view  $\psi_{-\mathbf{k}\downarrow}$ and $\psi_{\mathbf{k}\uparrow}$ as two copies of operators without the need of further duplication. Therefore, the inequivalent ground state labeled by the deformation parameter $\theta$ can be classified by the $q$-deformed Hopf algebra $h_{\mathbf{q}}(1|1)$, which is generated by the $q$-deformed operators
\begin{equation}\label{HABCS}
\{a_{\mathbf{k}},a^{\dagger}_{\mathbf{k}}\}=[2\mathcal{H}]_{q_{\mathbf{k}}},  [N, a_{\mathbf{k}}]=-a_{\mathbf{k}},[N, a^{\dagger}_{\mathbf{k}}]=a^{\dagger}_{\mathbf{k}}, [\mathcal{H},\bullet]=0,
\end{equation}
where $[x]_{q_{\mathbf{k}}}=\frac{q_{\mathbf{k}}^x-q_{\mathbf{k}}^{-x}}{q_{\mathbf{k}}-q_{\mathbf{k}}^{-1}}$ and $\mathcal{H}$ is the central operator characterizing the algebra. One can check that $[x]_q=[x]_{q^{-1}}$ and $[x]^*_q=[x]_{q^*}$. Since $2\mathcal{H}=1$ here, $[2\mathcal{H}]_{q_{\mathbf{k}}}=1$. Then, the Hopf algebra $h_{\mathbf{q}}(1|1)$ is equipped with $q$-deformed coproducts defined by
\begin{eqnarray}
& &\Delta (e^{i\frac{\tau}{2}}\psi_{q_{\mathbf{k}}})=q^{\frac{1}{2}}_{\mathbf{k}}e^{i\frac{\tau}{2}}\psi_{\mathbf{k}\uparrow}+q^{-\frac{1}{2}}_{\mathbf{k}}e^{i\frac{\tau}{2}}
\psi_{-\mathbf{k}\downarrow}, \nonumber \\
& &\Delta (e^{-i\frac{\tau}{2}} \psi^{\dagger}_{q_{\mathbf{k}}})=q^{\ast\frac{1}{2}}_{\mathbf{k}}e^{-i\frac{\tau}{2}}\psi^{\dagger}_{\mathbf{k}\uparrow}+q^{\ast-\frac{1}{2}}_{\mathbf{k}}e^{-i\frac{\tau}{2}}\psi^{\dagger}_{-\mathbf{k}\downarrow},\nonumber\\
& &\Delta \mathcal{H}=1,\quad \Delta N=N_{\mathbf{k}\uparrow}+N_{-\mathbf{k}\downarrow},
\end{eqnarray}
where we have assigned a phase factor $e^{i\frac{\tau}{2}}$ to the $\psi$-quantum for convenience. One example of the coproduct which operates on two modes is the addition of angular momenta of two particles as illustrated in Ref.~\cite{VitielloP98}.

The Bogoliubov transformation is in fact a linear combination of
the $q$-deformed copodruct operations. For fermions, we choose $q_{\mathbf{k}}=e^{2i\theta_{\mathbf{k}}}$. Hence $q_{\mathbf{k}}^*=q_{\mathbf{k}}^{-1}$ and $[x]_{q_{\mathbf{k}}}=[x]_{q_{\mathbf{k}}^{-1}}=[x]_{q_{\mathbf{k}}^*}$. We introduce the following operators
\begin{eqnarray}
A_{q(\theta_{\mathbf{k}})}&\equiv&\frac{\Delta (e^{i\frac{\tau}{2}}\psi_{q(\theta_{\mathbf{k}})})}{\sqrt{[2]_q}} \nonumber \\
&=&\frac{1}{\sqrt{[2]_q}}
(e^{i\theta_{\mathbf{k}}+i\frac{\tau}{2}}\psi_{\mathbf{k}\uparrow}+e^{-i\theta_{\mathbf{k}}+i\frac{\tau}{2}}\psi_{-\mathbf{k}\downarrow}),\nonumber\\
B_{q(\theta_{\mathbf{k}})}&\equiv&\frac{1}{\sqrt{[2]_q}}\frac{1}{i}\frac{\partial \Delta (e^{i\frac{\tau}{2}}\psi_{q(\theta_{\mathbf{k}})})}{\partial \theta_{\mathbf{k}}} \nonumber \\
&=&\frac{1}{\sqrt{[2]_q}}
(e^{i\theta_{\mathbf{k}}+i\frac{\tau}{2}}\psi_{\mathbf{k}\uparrow}-e^{-i\theta_{\mathbf{k}}+i\frac{\tau}{2}}\psi_{-\mathbf{k}\downarrow}).
\end{eqnarray}
A set of operators with canonical commutation relations is given by the combinations
\begin{eqnarray}
A(\theta_{\mathbf{k}})&=&\frac{\sqrt{[2]_q}}{2\sqrt{2}}(A_{q(\theta_{\mathbf{k}})}+A_{q(-\theta_{\mathbf{k}})}+iA^{\dagger}_{q(\theta_{\mathbf{k}})}
-iA^{\dagger}_{q(-\theta_{\mathbf{k}})}), \nonumber \\
B(\theta_{\mathbf{k}})&=&\frac{\sqrt{[2]_q}}{2\sqrt{2}}(B_{q(\theta_{\mathbf{k}})}+B_{q(-\theta_{\mathbf{k}})}-iB^{\dagger}_{q(\theta_{\mathbf{k}})}
+iB^{\dagger}_{q(-\theta_{\mathbf{k}})}). \nonumber \\
\end{eqnarray}
Hence, the Bogoliubov transformation (\ref{BT2}) can be written as
\begin{eqnarray}
\alpha_{\mathbf{k}}&=&\frac{e^{-i\frac{\tau}{2}}}{\sqrt{2}}\big[A(\theta_{\mathbf{k}})+B(\theta_{\mathbf{k}})\big], \nonumber \\
\beta_{-\mathbf{k}}&=&\frac{e^{-i\frac{\tau}{2}}}{\sqrt{2}}\big[A(\theta_{\mathbf{k}})-B(\theta_{\mathbf{k}})\big].
\end{eqnarray}
Therefore, the Bogoliubov transformations generating the inequivalent BCS ground states may be classified by the $q$-deformed Hopf algebra $\bigoplus_{\mathbf{k}}h_{q_{\mathbf{k}}}(1|1)$ if the amplitude of order parameter is perturbed while its phase is fixed.

\subsection{Fixed particle-number approach}
Before presenting analogous results for bosonic superfluid, we give a brief summary of the fixed particle-number approach to superconductivity. In the BCS ground state (\ref{GS0}), the numbers of Cooper pairs are mixed from the beginning. Ground states with fixed number of Coopers have been considered soon after the BCS theory were discovered~\cite{Blatt60,Bayman60}. Ref.~\cite{BarangerPR63} presented a detailed analysis of the fixed particle-number and BCS ground states and showed that the two approaches give the same equations of state in the thermodynamic limit.
However, while the superposition of different numbers of Cooper pairs in the BCS theory is limited to the grand canonical ensemble, the fixed particle-number approach has encountered  difficulties of properly normalized multi-particle correlations. Similar challenges remain in other fixed particle-number approaches. For instance, Ref.~\cite{LeggettBook} has to introduce a quantity $\langle N-2 | \psi_{{\bf k}\uparrow}\psi_{-{\bf k}\downarrow} |N\rangle$ connecting $N$-particle and $(N-2)$-particle states for constructing the Cooper-pair wavefunction.

Following Ref.~\cite{BarangerPR63}, the BCS ground state (\ref{GS0}) can be reformulated as
\begin{eqnarray}\label{GS10}
	|0(\theta)\rangle
	&=&Ce^{\sum_{\mathbf{k}}\phi_\mathbf{k}\psi^{\dagger}_{\mathbf{k}\uparrow}\psi^{\dagger}_{-\mathbf{k}\downarrow}}|0\rangle =C\sum_{n=0}^{+\infty}\frac{1}{n!}|\phi_n\rangle,
\end{eqnarray}
where $C=\prod_\mathbf{k}\sqrt{\frac{E_\mathbf{k}+\xi_\mathbf{k}}{2E_\mathbf{k}}}$, $\phi_\mathbf{k}=\frac{v^*_{\mathbf{k}}}{u_{\mathbf{k}}}$, and $|\phi_n\rangle=(\sum_{\mathbf{k}}\phi_\mathbf{k}\psi^{\dagger}_{\mathbf{k}\uparrow}\psi^{\dagger}_{-\mathbf{k}\downarrow})^n|0\rangle$. The state $|\phi_n\rangle$ has a fixed particle-number $2n$ corresponding to $n$ Cooper pairs, and it is normalized as $\langle \phi_n|\phi_m\rangle=F(2n)\delta_{nm}$. The function $F(2n)$ was defined and analyzed in Ref.\cite{BarangerPR63}. We will use the index $k$ to denote $(\mathbf{k},\uparrow)$ and $\bar{k}$ to denote the quantum numbers of the state paired with $k$, i.e., $(-\mathbf{k},\downarrow)$.
By defining $\phi_{k\bar{k}}=-\phi_{\bar{k}k}=\phi_\mathbf{k}$, the state $|\phi_n\rangle$ can be written as a diagonal form of Blatt's wave function (according to Ref.~\cite{BarangerPR63})
$
|\phi_{n}\rangle=(\sum_{\mathbf{k}}\phi_{k\bar{k}}\psi^\dagger_{k}\psi^\dagger_{\bar{k}})^n|0\rangle$.

Unlike the BCS ground state, the fixed particle-number ground state is not a coherent state, and it then does not saturate the Robertson-Schrodinger uncertainty relation. With the details summarized in Appendix~\ref{App:FixedN}, one can show that
\begin{eqnarray}
	\langle\phi_n|(\Delta S^x_{\mathbf{k}})^2|\phi_n\rangle\langle\phi_n|(\Delta S^y_{\mathbf{k}})^2|\phi_n\rangle>\frac{1}{4}|\langle\phi_n|[S^x_{\mathbf{k}},S^y_{\mathbf{k}}]|\phi_n\rangle|^2 \nonumber\\
\end{eqnarray}
whenever the pairing gap is finite. It follows the Robertson-Schrodinger uncertainty relation \eqref{eq:RSine} is not saturated. The reason why the BCS ground state saturates the relation \eqref{eq:RSine} while the fixed number wave function does not is because the uncertainty relation contains expectation values of particle-number changing operators like $S^{\pm}_\mathbf{k}$. For example,  $\langle\phi_n|S^+_{\mathbf{k}}|\phi_n\rangle=0$, but $\langle\phi_n|S^+_{\mathbf{k}}|\phi_{n-2}\rangle\neq0$. Therefore, when we calculate the expectation values of operators like $S^+_{\mathbf{k}}$ in the BCS ground state, mixed particle-number terms like $\langle\phi_n|S^+_{\mathbf{k}}|\phi_{n-2}\rangle$ must be included. As a consequence, the generalized uncertainty relation distinguishes the coherent-state structure of the two different (BCS and fixed particle-number) ground states.

Although the particle number of the wave function $\phi_n$ is fixed, it still possesses internal entanglement between the opposite spins at various momenta. Before we evaluate the entanglement entropy, the wave function must be normalized as discussed in Appendix~\ref{App:FixedN}. By casting the fixed particle-number ground state in a Schmidt decomposition, the entanglement entropy is
\begin{eqnarray}
	S_n=-\sum_{\mathbf{k};\{n\}}\frac{(n!)^2W_{\mathbf{k};\{n\}}}{C^2F(2n)}\ln \frac{(n!)^2W_{\mathbf{k};\{n\}}}{C^2F(2n)},
\end{eqnarray}
where $W_{\mathbf{k};\{n\}}=\prod_{\mathbf{k};\{n\}'}\big(\frac{E_{\mathbf{k}}-\xi_{\mathbf{k}}}{2E_{\mathbf{k}}}\big)^{n_{\mathbf{k}}}
\big(\frac{E_{\mathbf{k}}+\xi_{\mathbf{k}}}{2E_{\mathbf{k}}}\big)^{1-n_{\mathbf{k}}}$ and $F(2n)$ can be evaluated by using the relation  $|\phi_{k\bar{k}}|=\tan\theta_\mathbf{k}=\sqrt{\frac{E_{\mathbf{k}}-\xi_{\mathbf{k}}}{E_{\mathbf{k}}+\xi_{\mathbf{k}}}}$.
We remark that the entanglement entropy $S$ of the BCS ground state is not a simple sum of the entanglement entropy $S_n$ of the fixed particle-number ground states, i.e., $S\neq\sum_{n=0}^\infty S_n$ due to the normalizations in the two different approaches.

Finally, the Berry phase of the fixed particle-number wavefunction with $n$ Cooper pairs (equivalent to $2n$ fermions) is $\gamma_n=2n\pi$ as shown in Appendix~\ref{App:FixedN}. We remark that the Berry phase of the BCS ground state with averaged particle number $N$ is $\gamma=N\pi$. Thus, by choosing the particle number of the fixed particle-number ground state to be the same as the averaged particle number of the BCS ground state ($N=2n$), the two values of Berry phase agree. Indeed, Ref.~\cite{BarangerPR63} showed that the two ground states should lead to the same equations of state in the thermodynamic limit under this $N=2n$ condition.

\section{Bosonic BEC superfluids}\label{sec:BEC}
Similar investigations on the ground state of weakly interacting Bose gases will be presented here. The ground state should correspond to Bose-Einstein condensation (BEC)~\cite{London_book}. For convenience, we adopt the same notions from their BCS counterparts wherever applicable. It will be clear in a moment that the BCS and BEC ground states share many similar algebraic properties.

\subsection{BEC ground state}
Here the basic framework of a weakly interacting Bose gas is obtained by mostly following Ref.~\cite{Fetter_book}. The Hamiltonian is given by
\begin{eqnarray}
H=\sum_{\mathbf{k}}\frac{\mathbf{k}^2}{2m}a^{\dagger}_{\mathbf{k}}a_{\mathbf{k}}+\frac{g}{2V}\sum_{\mathbf{kp}}a^{\dagger}_{\mathbf{k}}a^{\dagger}_{-\mathbf{k}}a_{\mathbf{p}}a_{-\mathbf{p}}.
\end{eqnarray}
Here $a^{\dagger}$ and $a$ are boson creation and annihilation operators, $g$ is the coupling constant, and $V$ is the system volume. The ground state $|0(\theta)\rangle$ satisfies $\langle0(\theta)|a^{\dagger}_{\mathbf{0}}a_{\mathbf{0}}|0(\theta)\rangle=N_0$, where $N_0$ is the number of particles occupying the zero-momentum (lowest-energy) state. The BEC indicates a macroscopically occupied quantum state, so it is natural to replace the operators $a_{\mathbf{0}}$ and $a_{\mathbf{0}}^{\dagger}$ by $c$ numbers
\begin{eqnarray}
a_{\mathbf{0}}\rightarrow\sqrt{N_0}e^{-i\tau}, a_{\mathbf{0}}^{\dagger}\rightarrow \sqrt{N_0}e^{i\tau}.
\end{eqnarray}
Here $\tau$ is the phase of the condensate. The condensate wave function is the order parameter just like the BCS gap function. By using the U(1) symmetry of the Hamiltonian, the phase of the condensate is usually chosen to be zero. Here we will explore interesting physics when the order parameter are allowed to be tuned. The interaction part of the Hamiltonian can be classified according to the number of times $a_{\mathbf{0}}$ and $a^{\dagger}_{\mathbf{0}}$ appear. When $\frac{N-N_0}{N}\ll 1$, where $N$ is the total particle number, we can retain only terms of order $N^2_0$ and $N_0$ and neglect the interactions of particles out of the condensate. The interaction term of the Hamiltonian can be approximated by
\begin{eqnarray}\label{BECt1}
H_{\textrm{int}}&\approx&\frac{g}{2V}(a^{\dagger}_{\mathbf{0}})^2(a_{\mathbf{0}})^2+\frac{g}{2V}\sum'_{\mathbf{k}}[2(a^{\dagger}_{\mathbf{k}}a_{\mathbf{k}}a^{\dagger}_{\mathbf{0}}a_{\mathbf{0}}\nonumber\\
& &+a^{\dagger}_{-\mathbf{k}}a_{-\mathbf{k}}a^{\dagger}_{\mathbf{0}}a_{\mathbf{0}})+a^{\dagger}_{\mathbf{k}}a^{\dagger}_{-\mathbf{k}}a_{\mathbf{0}}a_{\mathbf{0}}+
a^{\dagger}_{\mathbf{0}}a^{\dagger}_{\mathbf{0}}a_{\mathbf{k}}a_{-\mathbf{k}}]\nonumber\\
&=&\frac{g}{2V}N^2_0+\frac{g}{2V}N_0\sum'_{\mathbf{k}}(2a^{\dagger}_{\mathbf{k}}a_{\mathbf{k}}+2a^{\dagger}_{-\mathbf{k}}a_{-\mathbf{k}}\nonumber\\
& &+e^{-2i\tau}a^{\dagger}_{\mathbf{k}}a^{\dagger}_{-\mathbf{k}}+e^{2i\tau}a_{\mathbf{k}}a_{-\mathbf{k}}),
\end{eqnarray}
where $\sum'_{\mathbf{k}}$ denotes the summation over $\mathbf{k}\neq\mathbf{0}$ only.
The particle number operator becomes
$\hat{N}=N_0+\frac{1}{2}\sum'_{\mathbf{k}}(a^{\dagger}_{\mathbf{k}}a_{\mathbf{k}}+a^{\dagger}_{-\mathbf{k}}a_{-\mathbf{k}})$.
If we consider the grand-canonical ensemble, the particle number is not conserved. Instead of introducing the chemical potential, we follow Ref.~\cite{Walecka} and consider $N=\langle \hat{N}\rangle$ as given and find the answer consistent with $N_0$.
After collecting the terms, the Hamiltonian becomes
$H_{\textrm{BEC}}=\frac{1}{2}Vgn^2+\frac{1}{2}\sum'_{\mathbf{k}}[(\epsilon_{\mathbf{k}}+ng)(a^{\dagger}_{\mathbf{k}}a_{\mathbf{k}}+a^{\dagger}_{-\mathbf{k}}a_{-\mathbf{k}})+ng
(e^{-2i\tau}a^{\dagger}_{\mathbf{k}}a^{\dagger}_{-\mathbf{k}}+e^{2i\tau}a_{\mathbf{k}}a_{-\mathbf{k}})]$,
where $n=\frac{N}{V}$ and $\epsilon_{\mathbf{k}}=\frac{\mathbf{k}^2}{2m}$. High order terms like $(\sum'_{\mathbf{k}}a^{\dagger}_{\mathbf{k}}a_{\mathbf{k}})^2$ have been neglected due to the assumption $\frac{N-N_0}{N}\ll 1$.

To diagonalize the Hamiltonian, we introduce the Bogoliubov transformation
\begin{eqnarray}
\alpha_{\mathbf{k}}=u_{\mathbf{k}}a_{\mathbf{k}}-v_{\mathbf{k}}a^{\dagger}_{-\mathbf{k}}.
\end{eqnarray}
Assuming $u_{\mathbf{k}}=u_{-\mathbf{k}}$, $v_{\mathbf{k}}=v_{-\mathbf{k}}$, we have
$a_{\mathbf{k}}=u^*_{\mathbf{k}}\alpha_{\mathbf{k}}-v_{\mathbf{k}}\alpha^{\dagger}_{-\mathbf{k}}$,
and the condition $|u_{\mathbf{k}}|^2-|v_{\mathbf{k}}|^2=1$ is imposed to ensure $[\alpha_{\mathbf{k}},\alpha^{\dagger}_{\mathbf{k}}]=1$. For simplicity, we choose $u_{\mathbf{k}}=|u_{\mathbf{k}}|$ and $v_{\mathbf{k}}=|v_{\mathbf{k}}|e^{-2i\tau}$.
To diagonalize the Hamiltonian, we impose the condition \cite{Walecka}
\begin{eqnarray}\label{dHC}
ng(|u_{\mathbf{k}}|^2+
|v_{\mathbf{k}}|^2)=2(\epsilon_{\mathbf{k}}+ng)|u_{\mathbf{k}}v_{\mathbf{k}}|.
\end{eqnarray}
Here $E_{\mathbf{k}}=\sqrt{\xi_{\mathbf{k}}^2-(ng)^2}$ with $\xi_{\mathbf{k}}=\epsilon_{\mathbf{k}}+ng$. The Hamiltonian is diagonalized as
\begin{eqnarray}\label{HBEC}
H_{\textrm{BEC}}&=&\frac{1}{2}\sum'_{\mathbf{k}}E_{\mathbf{k}}(\alpha^{\dagger}_{\mathbf{k}}\alpha_{\mathbf{k}}+
\alpha^{\dagger}_{-\mathbf{k}}\alpha_{-\mathbf{k}}) \nonumber \\
& &-\frac{1}{2}\sum'_{\mathbf{k}}(\epsilon_{\mathbf{k}}+ng-E_{\mathbf{k}})+\frac{1}{2}Vgn^2.
\end{eqnarray}
Since the energy dispersion of the quasi-particle $\alpha_{\mathbf{k}}$ is nonnegative, the ground state $|0(\theta)\rangle$ must be the Fock-space vacuum of the quasi-particle, i.e., $\alpha_{\mathbf{k}}|0(\theta)\rangle=0$ for all $\mathbf{k}\neq\mathbf{0}$. The number of particles with non-zero momentum in the ground state is given by
\begin{eqnarray}\label{BECN}
N-N_0&=&\sum'_{\mathbf{k}}\langle0(\theta)|a^{\dagger}_{\mathbf{k}}a_{\mathbf{k}}|0(\theta)\rangle=\sum'_{\mathbf{k}}|v_{\mathbf{k}}|^2.
\end{eqnarray}

Next, we reformulate the Bogoliubov transformation by introducing the relations $u_{\mathbf{k}}=\cosh\theta_{\mathbf{k}}$
 and $v_{\mathbf{k}}=-\sinh\theta_{\mathbf{k}}e^{2i\tau}$. Thus,
\begin{eqnarray}\label{BTB1}
\cosh2\theta_{\mathbf{k}}=\frac{\xi_{\mathbf{k}}}{E_{\mathbf{k}}},\quad \sinh2\theta_{\mathbf{k}}=\frac{ng}{E_{\mathbf{k}}}.
\end{eqnarray}
The Bogoliubov transformation can be written as
\begin{eqnarray}
\alpha_{\mathbf{k}}=\cosh\theta_{\mathbf{k}}a_{\mathbf{k}}+\sinh\theta_{\mathbf{k}}e^{-2i\tau}a^{\dagger}_{-\mathbf{k}}=e^Ga_{\mathbf{k}}e^{-G},
\end{eqnarray}
where the generator $G=\frac{1}{2}\sum'_{\mathbf{k}}\theta_{\mathbf{k}}(a_{\mathbf{k}}a_{-\mathbf{k}}e^{2i\tau}-a^{\dagger}_{\mathbf{k}}a^{\dagger}_{-\mathbf{k}}e^{-2i\tau})$ satisfies $(e^G)^{\dagger}=e^{-G}$. Just as its BCS counterpart, the ground state is obtained by the relation $|0(\theta)\rangle\equiv e^G|0\rangle$, where the non-interacting ground state $|0\rangle$ only contains condensed bosons with zero momentum. Since $a_{\mathbf{k}}|0\rangle=0$ if $\mathbf{k}\neq \mathbf{0}$, so $|0\rangle$ is the Fock-space vacuum of the $a_{\mathbf{k}}$-quantum with $\mathbf{k}\neq \mathbf{0}$.

It can be shown that the BEC ground state is an SU(1,1) generalized coherent squeeze state. We define  $S^-_{\mathbf{k}}=a_{\mathbf{k}}a_{-\mathbf{k}}$, $S^+_{\mathbf{k}}=a^{\dagger}_{\mathbf{k}}a^{\dagger}_{-\mathbf{k}}$, and  $S^z_{\mathbf{k}}=\frac{1}{2}(a^{\dagger}_{\mathbf{k}}a_{\mathbf{k}}+a_{-\mathbf{k}}a^{\dagger}_{-\mathbf{k}})=\frac{1}{2}(a^{\dagger}_{\mathbf{k}}a_{\mathbf{k}}+a^{\dagger}_{-\mathbf{k}}a_{-\mathbf{k}}+1)$, and it follows that  $[S^+_{\mathbf{k}},S^-_{\mathbf{k}}]=-2S^z_{\mathbf{k}}$ and $[S^z_{\mathbf{k}},S^{\pm}_{\mathbf{k}}]=\pm S^{\pm}_{\mathbf{k}}$. Therefore, $S^{\pm}_{\mathbf{k}}$ and $S^{z}_{\mathbf{k}}$ form an SU(1,1) algebra. It can be shown that $S^-_{\mathbf{k}}|0\rangle=0$ and $S^z_{\mathbf{k}}|0\rangle=\frac{1}{2}|0\rangle$. Let $w_{\mathbf{k}}=-\theta_{\mathbf{k}}e^{-2i\tau}$, then
$G=\frac{1}{2}\sum_{\mathbf{k}} '(w_{\mathbf{k}}S^+_{\mathbf{k}}-w^*_{\mathbf{k}}S^-_{\mathbf{k}})
=\sum_{\mathbf{k}}''(w_{\mathbf{k}}S^+_{\mathbf{k}}-w^*_{\mathbf{k}}S^-_{\mathbf{k}})$, where $\sum_{\mathbf{k}}''$ denotes the summation over $\mathbf{k}=(k_x,k_y,k_z)$ with $k_x>0$. A straightforward calculation (summarized in Appendix~\ref{appBEC}) shows that the Robertson-Schrodinger uncertainty relation, Eq.~\eqref{eq:RSine}, for bosons is satisfied with an equal sign.

By this convention $a_{\mathbf{k}}$ and $a_{-\mathbf{k}}$ are independent operators and every $S^{\pm}_{\mathbf{k}}$ only appears once in the $\mathbf{k}$-summation of $G$. Hence, we can directly apply the Baker-Campbell-Hausdorff disentangling formula to simplify the expression of the ground state
$|0(\theta)\rangle=e^{\sum_{\mathbf{k}}''\zeta_{\mathbf{k}}S^+_{\mathbf{k}}}e^{\sum_{\mathbf{k}}''\ln(1-|\zeta_{\mathbf{k}}|^2)S^z_{\mathbf{k}}}e^{-\sum_{\mathbf{k}}''\zeta^*_{\mathbf{k}}S^-_{\mathbf{k}}}|0\rangle
=e^{\sum_{\mathbf{k}}''\zeta_{\mathbf{k}}S^+_{\mathbf{k}}}e^{\frac{1}{2}\sum_{\mathbf{k}}''\ln(1-|\zeta_{\mathbf{k}}|^2)}|0\rangle$. We also define
\begin{equation}\label{eq:BECZeta} \zeta_{\mathbf{k}}=\zeta_{\mathbf{k}}(w_{\mathbf{k}})\equiv\frac{w_{\mathbf{k}}\tanh(|w_{\mathbf{k}}|)}{|w_{\mathbf{k}}|}=-\tanh(\theta_{\mathbf{k}})e^{-2i\tau}.
\end{equation}
By using  $\ln(1-|\zeta_{\mathbf{k}}|^2)=-2\ln\cosh\theta_{\mathbf{k}}$, the ground state becomes
\begin{eqnarray}\label{gBEC}
|0(\theta)\rangle&=&e^{-\sum_{\mathbf{k}}''\ln\cosh\theta_{\mathbf{k}}}e^{-\sum_{\mathbf{k}}''\tanh\theta_{\mathbf{k}}e^{-2i\tau}S^+_{\mathbf{k}}}|0\rangle \nonumber \\
&=&\prod''_{\mathbf{k}}\frac{1}{\cosh\theta_{\mathbf{k}}}e^{-\tanh\theta_{\mathbf{k}}e^{-2i\tau}a^{\dagger}_{\mathbf{k}}a^{\dagger}_{-\mathbf{k}}}|0\rangle.
\end{eqnarray}
A similar expression of the BEC ground state has been obtained by a different method~\cite{Ueda_book} with $\tau=0$, where it is called the Bogoliubov ground state.

\subsection{Berry phase of BEC ground state}
The Hamiltonian (\ref{HBEC}) also has a family of ground states which forms a U(1) manifold since the condensate can have an arbitrary phase. When a particular point on the manifold is chosen as the ground state, the U(1) symmetry is spontaneously broken. If the phase of the condensate is deformed along the U(1) manifold and back to its initial value, the system is transported along a closed loop $C$ parameterized by $\tau$ with $0\le\tau\le2\pi$ in the parameter space. In this way the system acquires a Berry phase.

The Berry phase for the SU(1,1) group has been discussed in Ref.~\cite{FHY88} although we believe there is a sign error in the derivation. Since the BEC ground state is an SU(1,1) generalized coherent state, we can apply a similar method to obtain its Berry phase as
\begin{eqnarray}
\gamma(C)=i\int_0^{2\pi}d\tau\langle 0(\theta(\tau))|\frac{d}{d\tau}|0(\theta(\tau))\rangle,
\end{eqnarray}
where
$|0(\theta(\tau))\rangle=e^{\frac{1}{2}\sum''_{\mathbf{k}}\ln(1-|\zeta_{\mathbf{k}}(\tau)|^2)}e^{\sum''_{\mathbf{k}}\zeta_{\mathbf{k}}(\tau)S^+_{\mathbf{k}}}|0\rangle$.
It can be verified that
$\langle0(\theta)|S^+_{\mathbf{k}}|0(\theta)\rangle=\frac{\zeta^*_{\mathbf{k}}}{1-|\zeta_{\mathbf{k}}|^2}$,
where $\zeta_{\mathbf{k}}$ is given by Eq.~(\ref{eq:BECZeta}).
Then, the Berry phase of the BEC ground state is
\begin{eqnarray}
\gamma(C)&=&i\int_0^{2\pi} d\tau \sum''_{\mathbf{k}}\left(-\frac{1}{2}\frac{\zeta^*_{\mathbf{k}}\frac{d\zeta_{\mathbf{k}}}{d\tau} +\zeta_{\mathbf{k}}\frac{d\zeta^*_{\mathbf{k}}}{d\tau}}{1-|\zeta_{\mathbf{k}}|^2} \right. \nonumber \\
& &\left. +\langle 0(\theta(\tau))|S^+_{\mathbf{k}}|0(\theta(\tau))\rangle\frac{d\zeta_{\mathbf{k}}}{d\tau}\right)\nonumber\\
&=&2\sum''_{\mathbf{k}}\int_0^{2\pi}d\tau\sinh^2\theta_{\mathbf{k}}(\tau)\nonumber\\
&=&2(N-N_0)\pi,
\end{eqnarray}
where the equations of state of the zero-temperature BEC superfluid and Eq.~(\ref{BECN}) have been applied in the last line.

$(N-N_0)$ corresponds to the bosons excited out of the condensate. One can trace such a dependence back to Eq.~\eqref{BECt1}, where the phases of the condensed particles cancel each other but the non-condensed particles accumulate phases from their interactions with the condensed particles. In contrast to the BCS case where the Berry phase depends on the total particle number (see Eq.(\ref{BPBCS})), in the BEC case the Berry phase is proportional only to the number of excited particles. The Berry phase may be viewed as an internal label differentiating different systems, and our study shows that the BCS and BEC ground states can be labeled differently according to their Berry phases. This has an important implication because a two-component Fermi gas can undergo a BCS-BEC crossover when the attractive interaction is increased \cite{Leggett}. However, there must be a topological change in the ground state because the Berry phase of the BCS ground state does not agree with the Berry phase of the BEC ground state.

Importantly, the calculation of Berry phase is different from the method discussed in Ref.~\cite{AndersonRMP} for obtaining the particle number using an imposed canonical relation between the particle number and the phase of the condensate wave function. Here we do not need to impose any canonical relation, and the Berry phase is the geometrical phase (to be distinguished from the dynamical phase) after the system is transported along a closed path in the parameter space.

We are also interested in the implication of the Berry phase in the parameter space. Similar to the BCS counterpart, it can be expressed as
\begin{eqnarray}
\gamma(C)&=&2\sum''_{\mathbf{k}}\int_0^{2\pi}d\tau\sinh^2\theta_{\mathbf{k}}\nonumber\\
&=&\sum''_{\mathbf{k}}\int_0^{2\pi}d\tau\frac{x_{\mathbf{k}}dy_{\mathbf{k}}-y_{\mathbf{k}}dx_{\mathbf{k}}}{E_{\mathbf{k}}(z_{\mathbf{k}}+E_{\mathbf{k}})},
\end{eqnarray}
where (see Eqs.(\ref{BTB1}))
\begin{eqnarray}
& &x_{\mathbf{k}}=E_{\mathbf{k}}\sinh(2\theta_{\mathbf{k}})\cos2\tau=ng\cos2\tau,\nonumber\\
& &y_{\mathbf{k}}=E_{\mathbf{k}}\sinh(2\theta_{\mathbf{k}})\sin2\tau=ng\sin2\tau,\nonumber\\
& &z_{\mathbf{k}}=E_{\mathbf{k}}\cosh(2\theta_{\mathbf{k}})=\xi_{\mathbf{k}}.
\end{eqnarray}
Note that $E^2_{\mathbf{k}}=z^2_{\mathbf{k}}-x^2_{\mathbf{k}}-y^2_{\mathbf{k}}$, hence the vector $\mathbf{x}_{\mathbf{k}}=x_{\mathbf{k}}\mathbf{i}+y_{\mathbf{k}}\mathbf{j}+z_{\mathbf{k}}\mathbf{k}$ lives in a hyperboloid space with the metric $g_{ij}=\textrm{diag}(-1,-1,1)$.
Therefore, the Berry phase is the line integral of the Berry connection
\begin{eqnarray}\mathbf{A}_\mathbf{k}=\frac{(-y_{\mathbf{k}},x_{\mathbf{k}},0)}{E_{\mathbf{k}}(z_{\mathbf{k}}+E_{\mathbf{k}})}
=\frac{(-ng\sin2\tau,ng\cos2\tau,0)}{E_{\mathbf{k}}(\xi_{\mathbf{k}}+E_{\mathbf{k}})},\end{eqnarray}
which in the hyperboloid coordinates becomes
\begin{eqnarray}\label{mnBEC2}
\mathbf{A}_\mathbf{k}=(0,0,\frac{\cosh2\theta_{\mathbf{k}}-1}{E_{\mathbf{k}}\sinh2\theta_{\mathbf{k}}})=(0,0,\frac{\xi_{\mathbf{k}}-E_{\mathbf{k}}}{E_{\mathbf{k}} ng}).
\end{eqnarray}
Thus, the Berry phase can be evaluated by $\gamma(C)=\sum''_{\mathbf{k}}\int_0^{2\pi}d\tau \mathbf{A}_\mathbf{k}\cdot d\mathbf{x}_{\mathbf{k}}$.
We remark that the Berry connection (or the effective vector potential) has no singularity in the
hyperboloid parameter space.

The Berry curvature, which behaves like an effective magnetic field, is
\begin{eqnarray}\label{MMBEC}
\mathbf{B}_{\mathbf{k}}=\nabla\times\mathbf{A}_{\mathbf{k}}=\frac{\mathbf{n}_{\mathbf{k}}}{E^2_{\mathbf{k}}},
\end{eqnarray}
where $\mathbf{n}_{\mathbf{k}}=(\sinh2\theta_{\mathbf{k}}\cos2\tau,\sinh2\theta_{\mathbf{k}}\sin2\tau,\cosh2\theta_{\mathbf{k}})$ and $E_{\mathbf{k}}$ is the radius in the parameter space.
The expressions \eqref{mnBEC2} and \eqref{MMBEC} indicate a magnetic monopole of one unit charge located at the origin of the parameter space.
Since the hyperboloid parameter space is non-compact, a magnetic monopole may have different properties from one in the Euclidean space. For example,
the curl operation is defined as  $\nabla\times\mathbf{A}_{\mathbf{k}}=\epsilon^{ijk}\partial_iA_{\mathbf{k}j}\vec{e}_k$ with $\vec{e}_{1,2,3}=\mathbf{i},\mathbf{j},\mathbf{k}$ and the antisymmetric tensor is defined as $\epsilon^{i_1i_2i_3}=g^{i_1j_1}g^{i_2j_2}g^{i_3j_3}\epsilon_{j_1j_2j_3}$. It can be shown that $\epsilon^{ijk}=\epsilon_{ijk}$ just as in the Euclidean space.
Finally, the Dirac quantization condition is also satisfied because each boson carries one electric charge ($Q^{\textrm{BEC}}_{\textrm{e}}=Q_e$) and the monopole carries one magnetic charge ($Q^{\textrm{BEC}}_{\textrm{m}}=Q_m$).

\subsection{Entanglement entropy of BEC ground state}
In the construction of the BEC ground state, $a_{\mathbf{k}}$ and $a_{-\mathbf{k}}$ are treated as independent operators by including only $k_x>0$ of $\mathbf{k}=(k_x,k_y,k_z)$ in the summations. One can consider a bipartite structure consisting of $a_{\mathbf{k}}$ and $a_{-\mathbf{k}}$ in the aforementioned domain. The BEC ground state cannot be factorized into a products of the $a_{\mathbf{k}}$ and $a_{-\mathbf{k}}$ modes, and we will show that it has the structure of an entangled state.

The BEC ground state~(\ref{gBEC}) also has a Schmidt decomposition
(for details, please refer to Appendix.\ref{appBEC})
\begin{align}\label{s2}
	|0(\theta)\rangle&=\prod''_{\mathbf{k}}\sum_{n=0}^{\infty}\frac{(-1)^n}{n!}\frac{\sinh^n\theta_{\mathbf{k}}}{\cosh^{n+1}\theta_{\mathbf{k}}}e^{-2in\tau}(a^{\dagger}_{\mathbf{k}})^n
	(a^{\dagger}_{-\mathbf{k}})^n|0\rangle \notag \\
	&=\sum_{n=0}^\infty e^{-2in\tau}\sum^{\prime\prime,\neq}_{\mathbf{k};\{n\}}\sqrt{W_{\mathbf{k};\{n\}}}\prod^{\prime\prime,\neq}_{\mathbf{k};\{n\}'}|n_{\mathbf{k}}\rangle\otimes|n_{-\mathbf{k}}\rangle,
\end{align}
where
$W_{\mathbf{k};\{n\}}=\prod^{\prime\prime,\neq}_{\mathbf{k};\{n\}'}\frac{\sinh^{2n_{\mathbf{k}}}\theta_{\mathbf{k}}}{\cosh^{2(1+n_{\mathbf{k}})}
\theta_{\mathbf{k}}}$, $|n_{\mathbf{k}}\rangle=\frac{(a^\dagger_{\mathbf{k}})^{n_{\mathbf{k}}}}{\sqrt{n_{\mathbf{k}}!}}|0\rangle$, $\sum^{\prime\prime,\neq}_{\mathbf{k};\{n\}}$ means the summation is taken over all possible momentum $\mathbf{k}$ which is different from one another only when $n_\mathbf{k}>0$ and $\sum''_\mathbf{k}n_\mathbf{k}=n$, and $\prod^{\prime\prime,\neq}_{\mathbf{k};\{n\}'}$ means the product is taken over all possible momentum which is different from one another, and satisfies the condition $\sum''_\mathbf{k}n_{\mathbf{k}}=n$ with $0\le n_{\mathbf{k}}\le n$. Any finite momentum $\mathbf{k}$ must have $k_{x}>0$ in the calculations, and this is why we add the superscript $''$ to the summation and product. The amplitude $W_{\mathbf{k};\{n\}}$ satisfies $0\le W_{\mathbf{k};\{n\}}\le 1$.
It can be shown that
\begin{eqnarray}
\sum_{n=0}^{\infty}\sum^{\prime\prime,\neq}_{\mathbf{k};\{n\}}W_{\mathbf{k};\{n\}}&=&\prod''_{\mathbf{k}}\frac{1}{\cosh^2\theta_{\mathbf{k}}}\sum_{n_{\mathbf{k}}=0}^{\infty}\tanh^{2n_{\mathbf{k}}}\theta_{\mathbf{k}}=1.
\end{eqnarray}
$W_{\mathbf{k};\{n\}}$ thus measures the probability of  $\prod^{\prime\prime,\neq}_{\mathbf{k};\{n\}'}(|n_{\mathbf{k}}\rangle\otimes|n_{-\mathbf{k}}\rangle)$ in the BEC ground state. 
Similar to the BCS case, the entanglement entropy of the BEC ground state is, by using Eq.~\eqref{E1},
\begin{eqnarray}\label{EBEC2}
S
&=&-\lim_{n\rightarrow1^+}\frac{d}{dn}\sum_{m=0}^{\infty}\sum^{\prime\prime,\neq}_{\mathbf{k};\{m\}}W^n_{\mathbf{k};\{m\}}\nonumber\\
&=&-\lim_{n\rightarrow1^+}\frac{d}{dn}\prod''_{\mathbf{k}}\frac{1}{\cosh^{2n}\theta_\mathbf{k}}\sum_{m_\mathbf{k}=0}^\infty\tanh^{2nm_\mathbf{k}}\theta_\mathbf{k}\nonumber\\
&=&\sum''_{\mathbf{k}}(|u_{\mathbf{k}}|^{2}\ln|u_{\mathbf{k}}|^{2}-|v_{\mathbf{k}}|^{2}\ln|v_{\mathbf{k}}|^{2})\nonumber\\
&=&\sum''_{\mathbf{k}}\Big(\frac{\xi_{\mathbf{k}}+E_{\mathbf{k}}}{2E_{\mathbf{k}}}\ln\frac{\xi_{\mathbf{k}}+E_{\mathbf{k}}}{2E_{\mathbf{k}}}-
\frac{\xi_{\mathbf{k}}-E_{\mathbf{k}}}{2E_{\mathbf{k}}}\ln\frac{\xi_{\mathbf{k}}-E_{\mathbf{k}}}{2E_{\mathbf{k}}}\Big).\nonumber\\
\end{eqnarray}
where Eq.~(\ref{BTB1}) has been applied.
The entanglement entropy can also be obtained by taking the vacuum expectation value of the entanglement entropy operator (see Appendix \ref{appBEC})
\begin{eqnarray}\label{s8}
	S_{\mathbf{k}}=-\sum''_{\mathbf{k}}(a^{\dagger}_{\mathbf{k}}a_{\mathbf{k}}\ln\sinh^2\theta_{\mathbf{k}}-a_{\mathbf{k}}a^{\dagger}_{\mathbf{k}}\ln\cosh^2\theta_{\mathbf{k}}).
\end{eqnarray}
A similar operator with $\mathbf{k}\leftrightarrow -\mathbf{k}$ gives the same result.

\subsection{Algebraic structure of BEC ground state}
When the phase of the condensate is kept constant while the parameter $\theta_{\mathbf{k}}$ is deformed externally, the BEC ground state can be written as $|0(\theta)\rangle=e^{\frac{1}{2}\sum'_{\mathbf{k}}\theta_{\mathbf{k}}\mathcal{G}_{\mathbf{k}}}|0\rangle$ with $\mathcal{G}_{\mathbf{k}}=a_{\mathbf{k}}a_{-\mathbf{k}}e^{2i\tau}-a^{\dagger}_{\mathbf{k}}a^{\dagger}_{-\mathbf{k}}e^{-2i\tau}$. The relation
$e^{\frac{1}{2}\sum'_{\mathbf{k}}\theta'_{\mathbf{k}}\mathcal{G}_{\mathbf{k}}}|0(\theta)\rangle=e^{\frac{1}{2}\sum'_{\mathbf{k}}(\theta_{\mathbf{k}}+
\theta'_{\mathbf{k}})\mathcal{G}_{\mathbf{k}}}|0\rangle=|0(\theta+\theta')\rangle$
shows that the Bogoliubov transformation relates different BEC ground states. Therefore, the infinitely many ground states are also classified by the algebraic property of the
Bogoliubov transformation. We further show that the algebraic properties follow the $q$-deformation of the Hopf algebra $\bigoplus_{\mathbf{k}}h_{q_{\mathbf{k}}}(1)$~\cite{VPLA98}, where $q_{\mathbf{k}}=e^{2\theta_{\mathbf{k}}}$. The algebra $h_{\mathbf{q}}(1)$ is generated by
\begin{equation}
[a_{\mathbf{k}},a^{\dagger}_{\mathbf{k}}]=[2\mathcal{H}]_{q_{\mathbf{k}}}, [N, a_{\mathbf{k}}]=-a_{\mathbf{k}},[N, a^{\dagger}_{\mathbf{k}}]=a^{\dagger}_{\mathbf{k}}, [\mathcal{H},\bullet]=0.
\end{equation}
Obviously, here $2\mathcal{H}=1$. The relations of this bosonic algebra are almost the same as those in the fermionic superalgebra (\ref{HABCS}), with the first anti-commutator replaced by the commutator. Thus the different algebras reflect the correct spin-statistics.

As we mentioned previously, $a_{\mathbf{k}}$ and $a_{-\mathbf{k}}$ are independent operators when $k_x>0$.  Therefore, they naturally form two copies of operators. To construct the Bogoliubov transformation, we also assign a phase factor to the $a$-quantum. Then the $q$-deformed coproducts associated with the $h_{\mathbf{q}}(1)$ algebra are defined by
\begin{eqnarray}
& &\Delta (e^{i\tau}a_{q_{\mathbf{k}}})=q^{\frac{1}{2}}_{\mathbf{k}}e^{i\tau}a_{\mathbf{k}}+q^{-\frac{1}{2}}_{\mathbf{k}}e^{i\tau}a_{-\mathbf{k}}, \nonumber \\
& &\Delta (e^{-i\tau}a ^{\dagger}_{q_{\mathbf{k}}})=q^{\ast\frac{1}{2}}_{\mathbf{k}}e^{-i\tau}a^{\dagger}_{\mathbf{k}}+q^{\ast-\frac{1}{2}}_{\mathbf{k}}e^{-i\tau}a^{\dagger}_{-\mathbf{k}},\nonumber\\
& &\Delta \mathcal{H}=1,\quad \Delta N=N_{\mathbf{k}}+N_{-\mathbf{k}}.
\end{eqnarray}
The Bogoliubov transformation is further found to be a linear combination of the $q$-deformed coproduct operations. We define $q_{\mathbf{k}}\equiv q(\theta_{\mathbf{k}})= e^{2\theta_{\mathbf{k}}}$ and consider the following operators
\begin{eqnarray}
& &A_{q(\theta_{\mathbf{k}})}\equiv \frac{\Delta a_{q(\theta_{\mathbf{k}})}}{\sqrt{[2]_q}}
=\frac{1}{\sqrt{[2]_q}}(e^{\theta_{\mathbf{k}}+i\tau}a_{\mathbf{k}}+e^{-\theta_{\mathbf{k}}+i\tau}a_{-\mathbf{k}}), \nonumber \\
& &B_{q(\theta_{\mathbf{k}})}\equiv \frac{1}{\sqrt{[2]_{q}}}\frac{\partial \Delta a_{q}}{\partial \theta_{\mathbf{k}}}
=\frac{1}{\sqrt{[2]_{q}}}(e^{\theta_{\mathbf{k}}+i\tau}a_{\mathbf{k}}-e^{-\theta_{\mathbf{k}}+i\tau}a_{-\mathbf{k}}). \nonumber \\
& &
\end{eqnarray}
A set of operators with canonical commutation relations
is given by the combinations
\begin{eqnarray}
A(\theta_{\mathbf{k}})&=&\frac{\sqrt{[2]_q}}{2\sqrt{2}}(A_{q(\theta_{\mathbf{k}})}+A_{q(-\theta_{\mathbf{k}})}+B^{\dagger}_{q(\theta_{\mathbf{k}})}
-B^{\dagger}_{q(-\theta_{\mathbf{k}})}) \nonumber \\
B(\theta_{\mathbf{k}})&=&\frac{\sqrt{[2]_q}}{2\sqrt{2}}(B_{q(\theta_{\mathbf{k}})}+B_{q(-\theta_{\mathbf{k}})}-A^{\dagger}_{q(\theta_{\mathbf{k}})}
+A^{\dagger}_{q(-\theta_{\mathbf{k}})}) \nonumber \\
& &
\end{eqnarray}
Finally, the Bogoliubov transformation can be written as
\begin{eqnarray}
\alpha_{\mathbf{k}}&=&\frac{1}{\sqrt{2}}\big[A(\theta_{\mathbf{k}})+B(\theta_{\mathbf{k}})\big]e^{-i\tau}, \nonumber \\
\alpha_{-\mathbf{k}}&=&\frac{1}{\sqrt{2}}\big[A(\theta_{\mathbf{k}})-B(\theta_{\mathbf{k}})\big]e^{-i\tau}.
\end{eqnarray}
Therefore, the BEC ground states may be classified by the $q$-deformed Hopf algebra $\bigoplus_{\mathbf{k}}h_{q_{\mathbf{k}}}(1)$.

\section{Conclusion}\label{sec:conclusion}
In conclusion, we have shown that the BCS ground state of a fermionic superfluid and the BEC ground state of a weakly interacting Bose gas are second-order nilpotent-SU(2) and SU(1,1) generalized coherent states, respectively. The two ground states minimize the Robertson-Schrodinger uncertainty relation and can be viewed as generalized squeezed states. When the ground states are adiabatically transported along a closed path in their parameter space, finite Berry phases emerge and the BCS and BEC ground states show different dependence of the Berry phase on the particle number. The Berry phase is associated with the effective magnetic flux of a monopole in the parameter space.

The BCS and BEC ground states naturally have bipartite structures allowing for entangled quantum correlations, and we present the entanglement entropy quantifying the internal entanglement. We also present a comparison between the fixed particle-number approach of fermionic superfluids with the BCS ground state approach.
The $q$-deformed Hopf algebra characterizes the Bogoliubov transformations for the two types of superfluid ground states, and the algebraic method may find future applications in other many-body systems.

\textit{Acknowledgment} -- H.G. thanks Prof. Dong Ruan for stimulating discussions and is supported by the National Natural Science Foundation of China (Grants No. 11674051).

\appendix
\section{Additional Results For BCS Ground State}\label{appBCS}
Here we show the details of $C=\langle0|0(\theta)\rangle=\langle0|e^G|0\rangle$ relating to the normalization of the ground state. By using $\frac{\partial \theta_{\mathbf{p}}}{\partial \theta_{\mathbf{k}}}=\delta_{\mathbf{p}\mathbf{k}}$, we have
\begin{eqnarray}
\frac{d C}{d \theta_{\mathbf{k}}}&=&\langle0|\frac{d G}{d \theta_{\mathbf{k}}}e^G|0\rangle\nonumber\\
&=&e^{i\tau}\langle0|e^{G}(\cos\theta_{\mathbf{k}}\psi_{\mathbf{k}\uparrow}+\sin\theta_{\mathbf{k}}e^{-i\tau}\psi^{\dagger}_{-\mathbf{k}\downarrow})\times \nonumber \\
& &(\cos\theta_{\mathbf{k}}\psi_{-\mathbf{k}\downarrow}-\sin\theta_{\mathbf{k}}e^{-i\tau}\psi^{\dagger}_{\mathbf{k}\uparrow})|0\rangle\nonumber\\
&=&-\cos\theta_{\mathbf{k}}\sin\theta_{\mathbf{k}}C+\sin^2\theta_{\mathbf{k}}\frac{d C}{d \theta_{\mathbf{k}}},
\end{eqnarray}
where we have used
$\frac{d C}{d \theta_{\mathbf{k}}}=\langle0|e^G\frac{d G}{d \theta_{\mathbf{k}}}|0\rangle=e^{-i\tau}\langle0|e^G\psi^{\dagger}_{\mathbf{k}\uparrow}\psi^{\dagger}_{-\mathbf{k}\downarrow}|0\rangle=-e^{-i\tau}\langle0|e^G\psi^{\dagger}_{-\mathbf{k}\downarrow}\psi^{\dagger}_{\mathbf{k}\uparrow}|0\rangle$.
Therefore, $\frac{d C}{d \theta_{\mathbf{k}}}=-C\tan\theta_{\mathbf{k}}.$
Since $C(\theta_{\mathbf{k}}=0,\textrm{for all $\mathbf{k}$})=1$, the solution is $C=\prod_{\mathbf{k}}\cos\theta_{\mathbf{k}}$.
Moreover, $C'=\frac{C}{\prod_{\mathbf{k}}u_{\mathbf{k}}}=1$.
This is consistent with
$\langle0(\theta)|0(\theta)\rangle=\langle 0 |\prod_{\mathbf{k}}(|u_{\mathbf{k}}|^2+|v_{\mathbf{k}}|^2)|0\rangle=1$.

It is known that all SU(2) generalized coherent states minimize the Robertson-Schrodinger inequality~\cite{CWCPL09}. Since the BCS ground state is a second-order nilpotent SU(2) generalized coherent state, it also satisfies this condition. One can show that
\begin{eqnarray}
\textrm{cov}(S^x_{\mathbf{k}},S^x_{\mathbf{k}})&=&\frac{1}{4}\big[1-(u_{\mathbf{k}}v_{\mathbf{k}}+u^*_{\mathbf{k}}v^*_{\mathbf{k}})^2\big], \nonumber \\
\textrm{cov}(S^y_{\mathbf{k}},S^y_{\mathbf{k}})&=&\frac{1}{4}\big[1+(u_{\mathbf{k}}v_{\mathbf{k}}-u^*_{\mathbf{k}}v^*_{\mathbf{k}})^2\big],\nonumber\\
\textrm{cov}(S^x_{\mathbf{k}},S^y_{\mathbf{k}})&=&-\frac{1}{4i}\big(u^2_{\mathbf{k}}v^2_{\mathbf{k}}-u^{\ast2}_{\mathbf{k}}v^{\ast2}_{\mathbf{k}}\big), \nonumber \\
\langle0(\theta)|[S^x_{\mathbf{k}},S^y_{\mathbf{k}}]|0(\theta)\rangle&=&-\frac{i}{2}(|u_{\mathbf{k}}|^2-|v_{\mathbf{k}}|^2).
\end{eqnarray}
Thus, Eq.~\eqref{eq:RSine} is satisfied with an equal sign.

Now we show the Schmidt decomposition (\ref{GS2}) of the BCS ground state. Eq.(\ref{GS1}) is expanded as
\begin{widetext}
\begin{eqnarray}\label{GS20}
& &|0(\theta)\rangle
=C\prod_{\mathbf{k}}\sum_{n=0}^1e^{-in\tau}\tan^n\theta_{\mathbf{k}}(\psi^{\dagger}_{\mathbf{k}\uparrow})^n(\psi^{\dagger}_{-\mathbf{k}\downarrow})^n|0\rangle\nonumber\\
&=&C\Big(1+\sum_{\mathbf{k}}\tan\theta_{\mathbf{k}}e^{-i\tau}\psi^{\dagger}_{\mathbf{k}\uparrow}\psi^{\dagger}_{-\mathbf{k}\downarrow}\sum_{\mathbf{k}_1\neq\mathbf{k}_2}\tanh\theta_{\mathbf{k}_1}\tanh\theta_{\mathbf{k}_2}e^{-2i\tau}\psi^{\dagger}_{\mathbf{k}_1\uparrow}
\psi^{\dagger}_{-\mathbf{k}_1\downarrow}\psi^{\dagger}_{\mathbf{k}_2\uparrow}\psi^{\dagger}_{-\mathbf{k}_2\downarrow}+\cdots\Big)|0\rangle\nonumber\\
&=&C\Big(|0\rangle\otimes|0\rangle+\sum_{\mathbf{k}}\tan\theta_{\mathbf{k}}e^{-i\tau}|1_{\mathbf{k}\uparrow}\rangle\otimes|1_{-\mathbf{k}\downarrow}\rangle+
\sum_{\mathbf{k}_1\neq\mathbf{k}_2}\tan\theta_{\mathbf{k}_1}\tan\theta_{\mathbf{k}_2}e^{-2i\tau}|1_{\mathbf{k}_1\uparrow}1_{\mathbf{k}_2\uparrow}\rangle\otimes|1_{-\mathbf{k}_1\downarrow}1_{-\mathbf{k}_2\downarrow}\rangle+\cdots\Big)|0\rangle\nonumber\\
&=&C\sum_{n=0}^\infty e^{-in\tau}\sum_{\mathbf{k}_i\neq\mathbf{k}_j;i=1,\cdots,n}\prod_{i=1}^n\Big(\tan\theta_{\mathbf{k}_i}|1_{\mathbf{k}_i\uparrow}\rangle\otimes|1_{-\mathbf{k}_i\downarrow}\rangle\Big),
\end{eqnarray}
\end{widetext}
where $\sum_{\mathbf{k}_i\neq\mathbf{k}_j;i=1,\cdots,n}$ means $\sum_{\mathbf{k}_1,\mathbf{k}_2,\cdots,\mathbf{k}_n}$ with $\mathbf{k}_i\neq\mathbf{k}_j$. We now absorb $C$ into the expression of the BCS ground state, we have
\begin{eqnarray}\label{GS21}
|0(\theta)\rangle&=&\sum_{n=0}^\infty \sum_{\mathbf{k}_i\neq\mathbf{k}_j;i=1,\cdots,n}\prod_{i=1}^n\Big(\sin\theta_{\mathbf{k}_i}|1_{\mathbf{k}_i\uparrow}\rangle\otimes
|1_{-\mathbf{k}_i\downarrow}\rangle\Big)\nonumber\\
& &\times e^{-in\tau}\prod_{\mathbf{k}'\neq\mathbf{k}_i}\cos\theta_{\mathbf{k}'}\nonumber\\
&=&\sum_{n=0}^\infty \sum_{\mathbf{k};\{n\}}\prod_{\mathbf{k};\{n\}'}\sin^{n_\mathbf{k}}\theta_\mathbf{k}\cos^{1-n_\mathbf{k}}\theta_\mathbf{k}\nonumber\\
& &\times(\psi^\dagger_{\mathbf{k}\uparrow})^{n_\mathbf{k}}(\psi^\dagger_{-\mathbf{k}\downarrow})^{n_\mathbf{k}}|0\rangle.
\end{eqnarray}
This is exactly Eq.(\ref{GS2}).
Here $\prod_{\mathbf{k};\{n\}'}$ means the product is taken over all possible $\mathbf{k}$ with the restriction $\sum_{\mathbf{k}}n_\mathbf{k}=n$ with $n_\mathbf{k}=0$ or $1$, while $\sum_{\mathbf{k};\{n\}}$ means the summation is taken over all possible $\mathbf{k}$ only when $n_\mathbf{k}=1$ and  $\sum_{\mathbf{k}}n_\mathbf{k}=n$. Since $\psi^\dagger_{\mathbf{k}_i\sigma}\psi^\dagger_{\mathbf{k}_j\sigma}=0$ if $\mathbf{k}_i=\mathbf{k}_j$, we don't need to impose the condition $\mathbf{k}_i\neq\mathbf{k}_j$ any more.

\section{Additional Results For BEC Ground State}\label{appBEC}
To verify the the Robertson-Schrodinger inequality is minimized by the BEC ground state, we utilize the following relations
\begin{eqnarray}
& &[G,S^+_{\mathbf{k}}]=-\sum_{\mathbf{p}} ''w^*_{\mathbf{p}}[S^-_{\mathbf{p}},S^+_{\mathbf{k}}]=-2w^*_{\mathbf{k}}S^z_{\mathbf{k}},\nonumber\\
& &[G,S^-_{\mathbf{k}}]=\sum_{\mathbf{p}}''w_{\mathbf{p}}[S^+_{\mathbf{p}},S^-_{\mathbf{k}}]=-2w_{\mathbf{k}}S^z_{\mathbf{k}},\nonumber\\
& &[G,S^z_{\mathbf{k}}]=\sum_{\mathbf{p}}''\Big(w_{\mathbf{p}}[S^+_{\mathbf{p}},S^z_{\mathbf{k}}]-w^*_{\mathbf{p}}[S^-_{\mathbf{p}},S^z_{\mathbf{k}}]\Big)\nonumber\\
& &\qquad\quad=-w_{\mathbf{k}}S^+_{\mathbf{k}}-w^*_{\mathbf{k}}S^-_{\mathbf{k}}.
\end{eqnarray}
These identities lead to
 \begin{eqnarray}
& &[G,w^*_{\mathbf{k}}S^-_{\mathbf{k}}+w_{\mathbf{k}}S^+_{\mathbf{k}}]=-4|w_{\mathbf{k}}|^2S^z_{\mathbf{k}}=-4\theta^2_{\mathbf{k}}S^z_{\mathbf{k}},  \nonumber \\
& &[G,w^*_{\mathbf{k}}S^-_{\mathbf{k}}-w_{\mathbf{k}}S^+_{\mathbf{k}}]=0.
 \end{eqnarray}
 Therefore, we have
\begin{eqnarray}
e^{-G}(w^*_{\mathbf{k}}S^-_{\mathbf{k}}+w_{\mathbf{k}}S^+_{\mathbf{k}})e^{G}&=&(w^*_{\mathbf{k}}S^-_{\mathbf{k}}+w_{\mathbf{k}}S^+_{\mathbf{k}})\cosh2\theta_{\mathbf{k}} \nonumber \\
& &+2\theta_{\mathbf{k}}S^z_{\mathbf{k}}\sinh2\theta_{\mathbf{k}}.   \nonumber \\
e^{-G}(w^*_{\mathbf{k}}S^-_{\mathbf{k}}-\tau_{\mathbf{k}}S^+_{\mathbf{k}})e^{G}&=&w^*_{\mathbf{k}}S^-_{\mathbf{k}}-w_{\mathbf{k}}S^+_{\mathbf{k}}.
\end{eqnarray}
After some algebra, the covariance coefficients are
\begin{eqnarray}
\textrm{cov}(S^x_{\mathbf{k}},S^x_{\mathbf{k}})&=&\frac{1}{4}\big(\cosh^22\theta_{\mathbf{k}}-\sinh^22\theta_{\mathbf{k}}\sin^22\tau\big),\\
\textrm{cov}(S^y_{\mathbf{k}},S^y_{\mathbf{k}})&=&\frac{1}{4}\big(\cosh^22\theta_{\mathbf{k}}-\sinh^22\theta_{\mathbf{k}}\cos^22\tau\big),\\
\textrm{cov}(S^x_{\mathbf{k}},S^y_{\mathbf{k}})&=&\frac{1}{8}\sinh^24\theta_{\mathbf{k}}\sin4\tau.
\end{eqnarray}
One can verify that Eq.~\eqref{eq:RSine} is satisfied.
Therefore, the Robertson-Schrodinger uncertainty relation \eqref{eq:RSine} associated with $S^x_{\mathbf{k}}$ and $S^y_{\mathbf{k}}$ for any momentum $\mathbf{k}$ is minimized by the BEC state. We note that the validity of the relation (\ref{eq:RSine}) actually does not require the condition (\ref{dHC}), under which the Hamiltonian \eqref{HBEC} is diagonalized.

Now we show the Schmidt decomposition of the BEC ground state. Eq.(\ref{gBEC}) is expanded as
\begin{widetext}
\begin{eqnarray}\label{s20}
|0(\theta)\rangle
&=&C\Big[1-\sum''_{\mathbf{k}}\tanh\theta_{\mathbf{k}}e^{-2i\tau}a^{\dagger}_{\mathbf{k}}a^{\dagger}_{-\mathbf{k}}\nonumber\\
&+&\big(\sum''_{\mathbf{k}}\tanh^2\theta_{\mathbf{k}}e^{-4i\tau}\frac{(a^{\dagger}_{\mathbf{k}})^2}{\sqrt{2!}}\frac{(a^{\dagger}_{-\mathbf{k}})^2}{\sqrt{2!}}
+\sum''_{\mathbf{k}_1\neq\mathbf{k}_2}\tanh\theta_{\mathbf{k}_1}\tanh\theta_{\mathbf{k}_2}e^{-4i\tau}a^{\dagger}_{\mathbf{k}_1}
a^{\dagger}_{-\mathbf{k}_1}a^{\dagger}_{\mathbf{k}_2}a^{\dagger}_{-\mathbf{k}_2}\big)+\cdots\Big]|0\rangle\nonumber\\
&=&C\Big[|0\rangle\otimes|0\rangle-\sum''_{\mathbf{k}}\tanh\theta_{\mathbf{k}}e^{-2i\tau}|1_{\mathbf{k}}\rangle\otimes|1_{-\mathbf{k}}\rangle\nonumber\\
&+&\big(\sum''_{\mathbf{k}}\tanh^2\theta_{\mathbf{k}}e^{-4i\tau}|2_{\mathbf{k}}\rangle\otimes|2_{\mathbf{k}}\rangle
+\sum''_{\mathbf{k}_1\neq\mathbf{k}_2}\tanh\theta_{\mathbf{k}_1}\tanh\theta_{\mathbf{k}_2}e^{-4i\tau}|1_{\mathbf{k}_1}1_{\mathbf{k}_2}\rangle\otimes|1_{-\mathbf{k}_1}1_{-\mathbf{k}_2}\rangle\big)+\cdots\Big]|0\rangle\nonumber\\
&=&C\sum_{n=0}^\infty e^{-2in\tau}\sum''_{\mathbf{k}_i\neq\mathbf{k}_j;\{n\}}\prod_{i;\{n\}}\Big(\tanh^{n_{\mathbf{k}_i}}\theta_{\mathbf{k}_i}|n_{\mathbf{k}_i}\rangle\otimes|n_{-\mathbf{k}_i}\rangle\Big),
\end{eqnarray}
\end{widetext}
where $C=\prod''_{\mathbf{k}}\frac{1}{\cosh\theta_{\mathbf{k}}}$, $\sum''_{\mathbf{k}_i\neq\mathbf{k}_j;\{n\}}$ means $\sum_{\mathbf{k}_1,\mathbf{k}_2,\cdots,\mathbf{k}_n}''$ with the restriction that $\mathbf{k}_i\neq\mathbf{k}_j$ for any $1\le i\neq j\le n$, and $\sum_in_{\mathbf{k}_i}=n$ with $1\le n_{\mathbf{k}_i}\le n$, $\prod_{i;\{n\}}$ means $\prod_{i=1}^n$ with the restriction that $\sum_in_{\mathbf{k}_i}=n$ with $1\le n_{\mathbf{k}_i}\le n$. Now we absorb the constant $C$ into the expression
\begin{eqnarray}\label{s3}
|0(\theta)\rangle&=&\sum_{n=0}^\infty \sum''_{\mathbf{k}_i\neq\mathbf{k}_j;\{n\}}\prod_{i;\{n\}}\Big(\frac{\sinh^{n_{\mathbf{k}_i}}\theta_{\mathbf{k}_i}}{\cosh^{1+n_{\mathbf{k}_i}}
\theta_{\mathbf{k}_i}}|n_{\mathbf{k}_i}\rangle\otimes|n_{-\mathbf{k}_i}\rangle\Big)\nonumber\\
&\times&\prod''_{\mathbf{k}'\neq\mathbf{k}_i}\frac{1}{\cosh\theta_{\mathbf{k}'}}.
\end{eqnarray}
Since
\begin{eqnarray}\frac{1}{\cosh\theta_{\mathbf{k}'}}=\frac{\sinh^{n_{\mathbf{k}'}}\theta_{\mathbf{k}'}}{\cosh^{1+n_{\mathbf{k}'}}
\theta_{\mathbf{k}'}}\textrm{ with }n_{\mathbf{k}'}=0,\end{eqnarray}then we can include the zero $n_\mathbf{k}$ into the solution of the equation $\sum''_\mathbf{k}n_{\mathbf{k}}=n$ with $1\le n_{\mathbf{k}_i}\le n$, the solution now satisfies $0\le n_{\mathbf{k}_i}\le n$.
Hence
\begin{eqnarray}
\prod''_{i;\{n\}}\frac{\sinh^{n_{\mathbf{k}_i}}\theta_{\mathbf{k}_i}}{\cosh^{1+n_{\mathbf{k}_i}}
\theta_{\mathbf{k}_i}}\prod''_{\mathbf{k}'\neq\mathbf{k}_i}\frac{1}{\cosh\theta_{\mathbf{k}'}}=\prod^{\prime\prime,\neq}_{\mathbf{k};\{n\}'}\frac{\sinh^{n_{\mathbf{k}}}\theta_{\mathbf{k}}}{\cosh^{1+n_{\mathbf{k}}}
\theta_{\mathbf{k}}}, \nonumber
\end{eqnarray}
where $\prod^{\prime\prime,\neq}_{\mathbf{k};\{n\}'}$ means the product is taken over all possible momentum which is different from one another, and satisfies the condition $\sum''_\mathbf{k}n_{\mathbf{k}}=n$ with $0\le n_{\mathbf{k}}\le n$. Then the ground state is now expanded as
\begin{widetext}
\begin{eqnarray}\label{s4}
|0(\theta)\rangle=\sum_{n=0}^\infty e^{-2in\tau}\sum^{\prime\prime,\neq}_{\mathbf{k};\{n\}}\prod^{\prime\prime,\neq}_{\mathbf{k};\{n\}'}\Big(\frac{\sinh^{n_{\mathbf{k}}}\theta_{\mathbf{k}}}{\cosh^{1+n_{\mathbf{k}}}
\theta_{\mathbf{k}}}\frac{(a^\dagger_{\mathbf{k}})^{n_{\mathbf{k}}}}{\sqrt{n_{\mathbf{k}}!}}\frac{(a^\dagger_{-\mathbf{k}})^{n_{\mathbf{k}}}}{\sqrt{n_{\mathbf{k}}!}}\Big)|0\rangle
,\end{eqnarray}
\end{widetext}
where $\sum^{\prime\prime,\neq}_{\mathbf{k};\{n\}}$ means the summation is taken over all possible momentum $\mathbf{k}$ which is different from one another, only when $n_\mathbf{k}>0$ and $\sum''_\mathbf{k}n_\mathbf{k}=n$.

\section{Fixed-Number Superconducting Ground State}\label{App:FixedN}
Here we summarize some key results from the fixed particle-number ground state. Using  Ref.~\cite{BarangerPR63} we found
\begin{eqnarray}
& &\textrm{cov}(S^x_{\mathbf{k}},S^x_{\mathbf{k}})=\langle\phi_n|(\Delta S^x_{\mathbf{k}})^2|\phi_n\rangle=\frac{1}{4},\nonumber\\
& &\textrm{cov}(S^y_{\mathbf{k}},S^y_{\mathbf{k}})=\langle\phi_n|(\Delta S^y_{\mathbf{k}})^2|\phi_n\rangle=\frac{1}{4},\nonumber\\
& &\textrm{cov}(S^x_{\mathbf{k}},S^y_{\mathbf{k}})=0,\nonumber\\
& &|\langle\phi_n|[S^x_{\mathbf{k}},S^y_{\mathbf{k}}]|\phi_n\rangle|=\frac{1}{2}||u_{\mathbf{k}}|^2-|v_{\mathbf{k}}|^2|. \end{eqnarray}
Thus, the Robertson-Schrodinger uncertainty relation is not saturated.

To evaluate the internal entanglement entropy, the wave function can be normalized and cast into a Schmidt decomposition given by
\begin{eqnarray}\label{GSn3}
\frac{1}{\sqrt{F(2n)}}|\phi_{n}\rangle=\frac{n!e^{-in\tau}}{C\sqrt{F(2n)}}\sum_{\mathbf{k};\{n\}}\sqrt{W_{\mathbf{k};\{n\}}}|n_{\mathbf{k}\uparrow}\rangle\otimes|n_{-\mathbf{k}\downarrow}\rangle ,
\nonumber\\\end{eqnarray}
where $C=\prod_\mathbf{k}\sqrt{\frac{E_\mathbf{k}+\xi_\mathbf{k}}{2E_\mathbf{k}}}$, $W_{\mathbf{k};\{n\}}=\prod_{\mathbf{k};\{n\}'}\big(\frac{E_{\mathbf{k}}-\xi_{\mathbf{k}}}{2E_{\mathbf{k}}}\big)^{n_{\mathbf{k}}}
\big(\frac{E_{\mathbf{k}}+\xi_{\mathbf{k}}}{2E_{\mathbf{k}}}\big)^{1-n_{\mathbf{k}}}$ and $F(2n)$ can be evaluated by using the relation  $|\phi_{k\bar{k}}|=\tan\theta_\mathbf{k}=\sqrt{\frac{E_{\mathbf{k}}-\xi_{\mathbf{k}}}{E_{\mathbf{k}}+\xi_{\mathbf{k}}}}$.

Finally, the Berry phase of the fixed particle-number wave function can be evaluated as follows. Since $\phi_{\mathbf{k}}=\tan\theta_\mathbf{k}e^{-i\tau}$, we have
\begin{widetext}
 \begin{eqnarray}
\gamma_n&=&i\int_0^{2\pi}d\tau\langle \phi_n(\tau)|\frac{d}{d\tau}|\phi_n(\tau)\rangle\nonumber\\
&=&\frac{i}{F(2n)}\int_0^{2\pi}d\tau\langle0|
(\sum_{\mathbf{k}}\tan\theta_\mathbf{k}e^{i\tau}\psi_{-\mathbf{k}\downarrow}\psi_{\mathbf{k}\uparrow})^n(-in)
(\sum_{\mathbf{k}'}\tan\theta_\mathbf{k'}e^{-i\tau}\psi^{\dagger}_{\mathbf{k}'\uparrow}\psi^{\dagger}_{-\mathbf{k}'\downarrow})^n|0\rangle\nonumber\\
&=&\frac{n}{F(2n)}\int_0^{2\pi}d\tau\langle0|
(\sum_{\mathbf{k}}|\phi_\mathbf{k}|\psi_{-\mathbf{k}\downarrow}\psi_{\mathbf{k}\uparrow})^n
(\sum_{\mathbf{k}'}|\phi_\mathbf{k}'|\psi^{\dagger}_{\mathbf{k}'\uparrow}\psi^{\dagger}_{-\mathbf{k}'\downarrow})^n|0\rangle\nonumber\\
&=&n\int_0^{2\pi}d\tau\nonumber\\
&=&2n\pi,
\end{eqnarray}
\end{widetext}
where we have used the definition of $F(2n)$ given by Eq. (21) of Ref.~\cite{BarangerPR63}.

\bibliographystyle{apsrev}

\end{document}